\documentclass{amsart}
\usepackage[english]{babel}
\usepackage{amsmath}
\usepackage{graphicx}
\usepackage[foot]{amsaddr}
\usepackage{xcolor}
\usepackage{subcaption}
\usepackage{float}
\usepackage{multirow}
\newcommand{\fig}[1]{Fig.~\ref{#1}}{\color{blue}}
\newcommand{\tab}[1]{Tab.~\ref{#1}}
\usepackage{comment}
\usepackage{stackengine,graphicx}
\usepackage{tikz}
\usepackage[top=1in, bottom=1.25in, left=1.25in, right=1.25in]{geometry}
\usepackage{hyperref}
    \hypersetup{
        colorlinks   = true,
        allcolors   =blue    
    }
\usepackage{tikz}
\usetikzlibrary{calc,fit,backgrounds}
\usetikzlibrary{shapes.geometric, arrows}
\usetikzlibrary{chains,
                fit,
                positioning,
                shapes}
\tikzstyle{startstop} = [rectangle, rounded corners, minimum width=10cm, minimum height=1cm,text centered, draw=black, fill=red!30]
\tikzstyle{io} = [trapezium, trapezium left angle=70, trapezium right angle=110, minimum width=2cm, minimum height=1cm, text centered, text width=2cm, draw=black, fill=blue!30]
\tikzstyle{io1} = [trapezium, trapezium left angle=70, trapezium right angle=110, minimum width=3cm, minimum height=1cm, text centered, text width=3cm, draw=black, fill=green!30]
\tikzstyle{process} = [rectangle, minimum width=1cm, minimum height=1cm, text centered, text width=2.5cm, draw=black, fill=orange!30]
\tikzstyle{decision} = [diamond, minimum width=3cm, minimum height=0.2cm, text centered,text width=1.2cm, draw=black, fill=green!30]
\tikzstyle{decision1} = [diamond, minimum width=2.8cm, minimum height=0.1cm,text width=1.8cm, draw=black, fill=green!30]
\tikzstyle{arrow} = [thick,->,>=stealth]
\tikzstyle{matheq} = [node distance=8.75cm, text width=21em, minimum width=1.5cm, minimum height=2em, text centered]
\tikzstyle{startstop1} = [rectangle, rounded corners, minimum width=5.5cm, minimum height=1cm,text centered, draw=black, fill=red!30]
\tikzstyle{io2} = [trapezium, trapezium left angle=70, trapezium right angle=110, minimum width=2cm, minimum height=1cm, text centered, text width=2cm, draw=black, fill=blue!30]
\tikzstyle{io1} = [trapezium, trapezium left angle=70, trapezium right angle=110, minimum width=3cm, minimum height=1cm, text centered, text width=3cm, draw=black, fill=green!30]
\tikzstyle{process1} = [rectangle, minimum width=1cm, minimum height=1.5cm, text centered, text width=2.5cm, draw=black, fill=orange!30]

\def\u{{\bm u}}
\def\v{{\bm v}}

\def\U{{\bm U}}

\def\0{\boldsymbol{0}}

\def\el {\nonumber }

\newcommand{\bm}[1]{\mbox{\boldmath{$#1$}}}

\begin{document}

\title[A non-intrusive data-driven ROM for parametrized CFD-DEM numerical simulatons]{A non-intrusive data-driven reduced order model for parametrized CFD-DEM numerical simulations}
\author{Arash Hajisharifi$^1$, Francesco Romanò$^1$, Michele Girfoglio$^1$, Andrea Beccari$^2$, Domenico Bonanni$^2$ and Gianluigi Rozza$^1$}
\address{$^1$ mathLab, Mathematics Area, SISSA, via Bonomea 265, I-34136 Trieste, Italy}
\address{$^2$ Dompé Farmaceutici SpA, EXSCALATE, Via Tommaso De Amicis, 95, I-80131 Napoli, Italy}

\begin{abstract}
The investigation of fluid-solid systems is very important in a lot of industrial processes. From a computational point of view, the simulation of such systems is very expensive, especially when a huge number of parametric configurations needs to be studied. In this context, we develop a non-intrusive data-driven reduced order model (ROM) built using the proper orthogonal decomposition with interpolation (PODI) method for Computational Fluid Dynamics (CFD) - Discrete Element Method (DEM) simulations. The main novelties of the proposed approach
rely in (i) the combination of ROM and FV methods, (ii) a numerical sensitivity analysis of the ROM accuracy with respect to the number of POD modes and to the cardinality of the training set and (iii) a parametric study with respect to the Stokes number.  We test our
ROM on the fluidized bed benchmark problem. 
The accuracy of the ROM is assessed
against results obtained with the FOM both for Eulerian (the fluid volume fraction) and Lagrangian (position and velocity of the particles) quantities. We also  discuss the efficiency of our ROM approach.
\end{abstract}

\maketitle


\section{Introduction}
Chemical engineering processes involve several complex phenomena such as thermochemical reactions and multiphase flows. For a long time experimental tests have been considered as the main source for the comprehension of fluid-solid systems  and the design of novel industrial infrastructure. However, it is well known that reliable experimental measurements are typically very hard to be obtained. This is because the environment is very dangerous as it is characterized by high temperature and pressure as well as the presence of toxic substances. Another problem refers to  the high cost of the measurement instruments. In this scenario numerical simulations represent a very useful tool which is acquiring more and more importance.  They are actually used to complement the measurements but, in perspective, they could also provide a valid support to the planning of the experimental tests as well as the designing of industrial prototypes with a significant improvement of the process efficiency. 

In this work we deal with the Computational Fluid Dynamics (CFD) - Discrete Element Method (DEM) approach. It is a Eulerian-Lagrangian technique used for the simulation of
systems involving the interaction between a fluid flow and solid particles \cite{noro, hajisharifi2021particle, hajisharifi2022interface, review, clayton, heat}. 
One of the major drawbacks of the CFD-DEM technique is related to its high computational cost that strongly limits 
the number of configurations that could be simulated. 
Although thanks to the recent  improvement of high performance computing infrastructures and the introduction of the coarse-grained modelling \cite{di2021coarse}
the CFD-DEM approach is becoming affordable, we are still far from its extensive use in the industrial practise where often it is required a huge number of simulations at the aim to investigate different physical and/or geometrical configurations.

In this context, Reduced Order Models (ROMs) (see, e.g., \cite{hesthaven2016certified, quarteroni2015reduced, benner2015survey, benner2017model, benner2021system, bader}) 
could be proposed as a tool able to increase the efficiency of the CFD-DEM simulations in a parametric fashion without a significant loss of accuracy. 
The basic idea on which ROM is based is that the parametric dependence of the problem at hand has an intrinsic dimension which is much
lower than the number of degrees of freedom of the discretized system. In order
to reach this dimensionality reduction, firstly a database
of solutions is collected by solving the original high fidelity model (hereafter referred as Full Order Model (FOM)) for different physical
and/or geometrical configurations (\emph{offline} phase). Then, the information obtained during
the offline phase is used to compute the solution for new values of the parameters
in a short amount of time (\emph{online} phase). 

ROMs enable both intrusive and non-intrusive approaches, based on the
nature of the solver that is used. In the former case, one has access to the source
code of the FOM solver and this allows to directly work on the equations of the
original problem. In the latter one, one relies only on the solutions, without
requiring information about the physical system and the equations describing
it; therefore, it could be applied when one refers to closed source solvers that are widely used by the industrial companies. In general intrusive ROMs are more accurate than the data-driven ones as they are physics-based. On the other hand, the efficiency of intrusive ROMs is rather limited when dealing with non-linear equations where the affine decomposition assumption does
not hold true. So, for applications where a high speed-up is required, data-driven ROMs are to be preferred. 


At the best of our knowledge, the literature about the development of ROMs for CFD-DEM simulations is still rather poor and needs to be enriched. Preliminary numerical results, based on a POD-Galerkin approach, can be found in \cite{palacios, yuan2005reduced}. 
On the other hand, some ROMs for the DEM formulation only are addressed in 
\cite{wallin, fani}. 
Recently, an important step forward has been made in \cite{shuo, shuo2} 
where the authors have proposed a non intrusive data-driven ROM framework based on the so-called PODI (Proper Orthogonal Decomposition with Interpolation) approach both for Eulerian and Lagrangian variables: the POD is employed for the construction of reduced basis space whilst an interpolation procedure based on  Radial Basis Functions (RBFs) is used for the evaluation of the reduced coefficients. Their full order solver is based on a Finite Difference (FD) method coupled with an immersed boundary approach \cite{mori}. However in \cite{shuo, shuo2, mori} the investigation is limited to the time reconstruction of the fields. 
Indeed, the present contribution aims to corroborate and extend the investigation reported in \cite{shuo, shuo2, mori} 
by introducing the following novelties: 
\begin{itemize}

\item The use of the Finite Volume (FV) method for the space discretization. The FD method adopted in \cite{shuo, shuo2, mori} could exhibit several limitations when complex geometries should be addressed due to the necessity to work with structured meshes. Moreover, we highlight that many commercial (e.g. Ansys Fluent and STAR-CCM+) and academic (e.g. OpenFOAM) codes are based on FV methods. Thus, the combination of ROM and FV methods is appealing for practical applications. 
\item An extensive sensitivity analysis of the ROM error at varying of the number of POD modes (i.e. of the system's energy retained). In \cite{shuo, shuo2, mori} the authors deeply investigated the convergence of the Lanczos based POD (LPOD) method employed to generate a set of reduced basis. Moreover, they derived some theoretical bounds for the error. However, at the best of our knowledge, an exhaustive numerical investigation is missing. In addition, we also perform a sensitivity analysis of the ROM error at varying of the number of FOM snapshots used to train the ROM.
\item A preliminary parametric study of the Eulerian phase with respect to the Stokes number, which is 
a crucial parameter of the model. 
In fact, as already was mentioned, in \cite{shuo, shuo2} only the time reconstruction is considered 
but in the industrial practise the solution typically depends on a wide range of parameters. Therefore, an efficient parametric ROM is needed to be developed in order to move gradually towards real-world applications. 
Within this parametric investigation, we use two variants of the PODI approach: the global PODI and the local PODI. For the former we use global POD basis computed by time-dependent FOM snapshots associated to sample points in the parameter space, for the latter a POD basis is computed for each parameter in the training set and the basis functions for new parameter values are found via RBFs interpolation of the basis functions associated to the training set. This is because the global PODI, as we will see, is not able to provide reliable results. So we underline the difficulty to deal with a parametric ROM in the context of CFD-DEM modelling that represents an open challenge. 


\end{itemize}

The work is organized as follows. In Sec. \ref{sec:FOM} a brief description of the CFD-DEM model (i.e. our FOM), including some relevant insights about its numerical discretization, is presented. Then in Sec. \ref{sec:ROM} the ROM approach is described and the achieved results are introduced and discussed in Sec. \ref{sec:res}. Finally in Sec. \ref{sec:conc} conclusions are drawn and future perspectives are envisioned.


\section{The full order model}\label{sec:FOM}
The CFD-DEM model is based on an Eulerian approach for the fluid phase and a Lagrangian approach for the solid phase \cite{Xu, tsu}. In Sec. \ref{sec:fluid} we describe the governing equations of the fluid phase whilst in Sec. \ref{sec:solid} we present the governing equations of the solid phase. 


\subsection{Governing equations for the fluid-phase flow}\label{sec:fluid}

Let $\Omega$ be a fixed spatial domain and $(t_0, T]$ a time interval of interest. 
Then the volume-averaged Navier-Stokes equations (NSE) read \cite{clayton}: 
\begin{equation}
\centering
\frac{\partial \epsilon}{\partial t} + \nabla \cdot ( \epsilon \u ) = 0 \hspace{6cm} \text{in} \ \Omega \ \times \ (t_0, T],
\label{continuity}
\end{equation}
\begin{equation}
\frac{\partial (\epsilon \u)}{\partial t} + \nabla \cdot ( \epsilon \u \otimes \u) = - \nabla P - S_p + \nabla \cdot (\epsilon \bm{\tau}) + \epsilon \textbf{g} \hspace{1cm} \text{in} \ \Omega \ \times  \ (t_0, T],
\label{momentu}
\end{equation}
\\
endowed with proper boundary conditions 
and the initial data $\u = \u_0$ and $\epsilon = \epsilon_0$ in $\Omega \times\{t_0\}$.
In addition,  $\u$ is the fluid velocity, $\epsilon$ is the fluid volume fraction, $\bm g$ is the gravity, $P$ is the modified pressure (i.e. $p/ \rho_f$ where $p$ is the pressure and $\rho_f$ is the fluid density). 
The viscous stress tensor $\bm{\tau}$ is computed as: \\ 
\begin{equation}
\bm{\tau} = \bm{\tau}_1 + \bm{\tau}_2 = \nu_f \left( \nabla\u + \nabla\u^T \right)- \dfrac{2}{3} \nu_f \left(\nabla \cdot \u \right) \bm{I},
\label{tau}
\end{equation}
where $\nu_f$ = $\mu_f / \rho_f$ is the kinematic viscosity and $\bm{I}$ is the identity matrix.

We partition the computational domain $\Omega$ into cells or control volumes $\Omega_i$, with
$i = 1, \dots, N_c$, where $N_c$ is the total number of cells in the mesh. 
The fluid volume fraction $\epsilon_{i}$, i.e. the volume
fraction occupied by the fluid in the cell $i$, is defined as follows \cite{weller} 
\begin{equation}
\epsilon_{i} = 1 - \frac {\sum_{j=1}^{n_p} \widetilde{\Omega}_j}{\Omega_{i}},
\label{eps}
\end{equation}
where  $n_p$ and $\widetilde{\Omega}_j$ are the number of particles located in the cell $i$ and the volume of the particle $j$, respectively. 
The coupling between the fluid phase and the particles is enforced via the source term $S_p$ in the momentum equation \eqref{momentu}. In a discrete sense, the source term acting in the cell $i$, $S_{p,i}$, is computed as \cite{zhu2007discrete} 
\begin{equation}
S_{p,i} = \frac{\sum_{j=1}^{n_p} ( \mathbf{F}_{d,j} + \mathbf{F}_{\nabla p,j})}{\rho_f{\Omega}_{i}},
\label{source}
\end{equation}
where 
\begin{equation}
\mathbf{F}_{d,j} = \dfrac{\widetilde{\Omega}_j\beta}{\widetilde{\epsilon}_i}\left(\u_i - \bm{\widetilde{u}_j} \right), \quad \mathbf{F}_{\nabla p,j} = -\widetilde{\Omega}_j \nabla p,
\label{drag}
\end{equation}
are the drag force and pressure gradient force, $\widetilde{\u}_j$ is the velocity of the particle $j$ and $\widetilde{\epsilon_i} = 1 - \epsilon_i$ is the solid volume fraction. In addition, $\beta$ is the inter-phase momentum exchange cofficient that needs to be properly tuned. In this work we adopt the empirical correlations provided in \cite{ergun, wen}. 


For further details, 
we refer the reader to \cite{review, fernandes}. 

\subsubsection{Numerical discretization} 

Let us start with the time discretization.
Let $\Delta t \in \mathbb{R}$, $t^n = t_0 + n \Delta t$, with $n = 0, \dots, N_T$ and $T = t_0 + N_T \Delta t$. We denote by $f^n$ the approximation of a generic quantity $f$ at
the time $t^n$. 
Problem \eqref{continuity}-\eqref{momentu} discretized in time adopting a first-order implicit Euler scheme and a segregated algorithm reads: given $\epsilon_{0}$ and $\u_0$ for $n \geq 0$

\begin{itemize}
    \item Find the fluid volume fraction $\epsilon^{n+1}$ such that

\begin{equation}
\centering
\frac{\epsilon^{n+1} - \epsilon^{n}}{\Delta t} + \nabla \cdot ( \epsilon^{n} \u^{n} ) = 0, 
\label{continuity_time}
\end{equation}

\item 
Find the fluid flow velocity $\u^{n+1}$ such that
\begin{equation}
\frac{\epsilon^{n+1} \u^{n+1}}{\Delta t} + \nabla \cdot ( \epsilon ^{n+1} \u^{n} \otimes \u^{n+1}) + \nabla P^{n+1}  - \nabla \cdot (\epsilon^{n+1}  \bm{\tau}_1^{n+1}) - \nabla \cdot (\epsilon^{n+1}  \bm{\tau}_2^{n}) = \mathbf{b}^{n} + \epsilon^{n+1} \bm{g}, 
\label{momentum}
\end{equation}
where $\mathbf{b}^{n} = \epsilon^n \u^n/\Delta t - S_p^n$. 
\end{itemize}




For the space discretization, we adopt the Finite Volume (FV) approximation that is derived directly
from the integral form of the governing equations. 
Let $\bm{A}_k$ be the surface vector of each face of the control volume, with $k = 1, \dots, M$.

Let us denote with $\u_i$, $\epsilon_i$ and ${\bm b}_i$ the average velocity, the fluid volume fraction and the source term in control volume $\Omega_i$,  respectively.
Moreover, we denote with $\u_{i,k}$, $\epsilon_{i,k}$, $P_{i,k}$ and $\bm{\tau}_{i,k}$ the velocity, the fluid volume fraction, the pressure and the stress tensor
associated to the centroid of face $k$ normalized by the volume of $\Omega_i$.
Thus, the FV formulation of eqs. \eqref{continuity_time}-\eqref{momentum} for each volume $\Omega_i$ is given by: 
\begin{align}
\frac{1}{\Delta t} (\epsilon_i^{n+1} - \epsilon_i^{n})\, + \sum_j^{} \varphi^n_k \epsilon_{i,k}^n = 0, 
\end{align}
\begin{align}
\frac{1}{\Delta t}\ \epsilon_i^{n+1} \u^{n+1}_i &+ \sum_k^{} \varphi_k ^n\epsilon^{n+1}_{i,k} \u^{n+1}_{i,k} + \sum_k^{} P^{n+1}_{i,k} \textbf{A}_k - \sum_k^{} \epsilon^{n+1}_{i,k} \bm{\tau}^{n+1}_{1i,k} \cdot \textbf{A}_k - \sum_k^{} \epsilon^{n+1}_{i,k} \bm{\tau}^{n}_{2i,k} \cdot \textbf{A}_k= {\bm b}^{n}_i + \epsilon^{n+1}_i \bm{g},
\end{align}
where $\varphi_k^n = \u_k^n \cdot A_k$ is the convective flux associated to $\u$ through the \textit{k}-th surface of the control volume $\Omega_i$.
To deal with the pressure-velocity coupling, we choose a partitioned approach. In
particular a PIMPLE algorithm \cite{pimple} is used, it consists into the
combination of a SIMPLE \cite{simple} and PISO \cite{issa} procedure. For the implementation of the numerical scheme described in this section we chose the finite
volume C++ library OpenFOAM\textsuperscript{\textregistered}\cite{weller}. 


\subsection{Governing equations for the particle system}\label{sec:solid}

DEM solves the particle flow at the particle level \cite{review, molin, fernandes}, i.e. it calculates the trajectory of each particle considering the effect of other particles, walls or other problem-specific forces.
 \\
The equations describing the dynamics of particles are derived by the second Newton’s law for translation and rotation:
\begin{equation}
m_j \frac{d\widetilde{\u}_j}{dt} =  \sum_{m=1}^{n_j^c} \bm{F}_{jm}^c +  \bm{F}_j^f + m_j \bm g ,\\
\label{translation}
\end{equation}
\begin{equation}
I_j \frac{d\mathbf \omega_j}{dt} = \sum_{m=1}^{n_j^c} \bm{M}_{jm}^c, 
\label{rotation}
\end{equation}
endowed with initial data $\widetilde{\u}_j = \widetilde{\u}_{j,0}$ and $\omega_j = \omega_{j,0}$ for $j = 1, \dots, n_p$. In eqs. \eqref{translation}-\eqref{rotation} $m_j$ and $I_j$ are the mass and the moment of inertia of the particle $j$
\begin{equation}
m_j = \rho_p\dfrac{\pi d_p^3}{6} , \quad I_j = \dfrac{m_j d_p^3}{6},
\end{equation}
with $\rho_p$ and $d_p$ the density and the diameter of the particles, 
 $\mathbf \omega_j$ denotes the angular velocity of particle $j$,  $\bm{F}_{jm}^c$ and $\bm{M}_{jm}^c$ are the contact force and torque acting on particle $j$ by its $m$ contacts, either with a particle or a wall (whose explicit formulation can be found, e.g.,  in \cite{fernandes}), 
$n_j^c$ is the number of total contacts for particle $j$ and 
$\bm{F}_j^f = \bm{F}_{d,j} + \bm{F}_{\nabla p, j}$ is the particle-fluid interaction force acting on particle $j$ (see eqs. \eqref{source}-\eqref{drag}).
It should be noted that in this work 
non-contact forces are not taken into account.

We introduce the Stokes number $Stk$ which characterizes the behavior of particles suspended in a fluid flow:
\begin{equation}
    Stk = \frac{\tau_s}{\tau_f},
\label{Stk}
\end{equation}
where the carrier fluid characteristic time $\tau_f$ is defined as follows:
\begin{equation}
   \tau_f = \frac{L_r}{U_r},
\label{Stk2}
\end{equation}
with $L_r$ and $U_r$ the reference fluid flow length and velocity, respectively.  Concerning the particle relaxation time $\tau_p$, we have \cite{ hajisharifi2022interface, marchioli2002mechanisms, sheidani2022study}
\begin{equation}
    \tau_p = \frac{\rho_p \: d_p}{18 \: \mu_f}.
\label{Stk3}
\end{equation}

By adopting a first-order Euler scheme, the discretized form of eqs. \eqref{translation}-\eqref{rotation}, given $\widetilde{\u}_{j,0}$ and $\omega_{j,0}$, for $n \geq 0$ yields:

\begin{equation}
m_j \dfrac{\widetilde{\u}_j^{n+1}}{\Delta t} = m_j \bm g + \widetilde{\bm{b}}_t^n, \\ 
\label{translation_disc}
\end{equation}
\begin{equation}
I_j\dfrac{\omega_j^{n+1}}{\Delta t}  = \widetilde{\bm{b}}_r^n, 
\label{rotation_disc}
\end{equation}
where $\widetilde{\bm{b}}_t^n = m_j \dfrac{\widetilde{\u}_j^{n}}{\Delta t} + \left(\sum_{m=1}^{n_j^c} \bm{F}_{jm}^c +  \bm{F}_j^f\right)^n$ and $\widetilde{\bm{b}}_r^n = I_j\dfrac{\omega_j^{n}}{\Delta t} + \left(\sum_{m=1}^{n_j^c} \bm{M}_{jm}^c\right)^n$.

\section{The reduced order model}\label{sec:ROM}
We assume that any Eulerian (Lagrangian) variable can be approximated as a linear combination of a certain number of basis functions depending on the space $\bm{x}$ (label $l$ identifying the particles) only, multiplied by scalar coefficients that depend on the time and/or parameters of the problem at hand which can be physical or geometrical. 

We are going to reconstruct the time evolution of the system and to consider the Stokes number $Stk$ (see eq. \eqref{Stk}) as a parameter. 
For what concerns the Eulerian phase, we are interested in the reconstruction of the fluid volume fraction $\epsilon$ whilst for the Lagrangian phase we consider the position $\widetilde{\bm{x}}$
and the velocity $\widetilde{\u}$ of the particles.
Hence,
the variables $(\epsilon, \widetilde{\bm{x}},  \widetilde{\bm{u}})$ 
are approximated by the reduced ones $(\epsilon_r, \widetilde{\bm{x}}_r,  \widetilde{\bm{u}}_r)$ as follows 

\begin{align}
\epsilon \approx \epsilon_r = \sum_{i=1}^{N_{\epsilon}^r} \bm{\alpha}_i(t, \bm{\pi}) {\varphi}_i(\bm{x}), \quad 
\bm{\widetilde{x}} \approx \bm{\widetilde{x}}_r = \sum_{i=1}^{N_{\widetilde{x}}^r} \bm{\beta}_i(t, \bm{\pi}) \bm{\xi}_i(l), \quad 
\bm{\widetilde{u}} \approx \bm{\widetilde{u}}_r = \sum_{i=1}^{N_{\widetilde{u}}^r} \bm{\gamma}_i(t, \bm{\pi}) \bm{\zeta}_i(l).\label{eq:ROM_1} 
\end{align}

In \eqref{eq:ROM_1}, $N_{\Phi}^r$ denotes the cardinality of a reduced basis for the space field $\Phi$ = $\{\epsilon, \widetilde{\bm{x}},  \widetilde{\bm{u}}\}$ and $\bm{\pi}$ is the parameter vector (that for the problem at hand it coincides with the Stokes number $Stk$). 

In this work we employ the proper orthogonal decomposition with interpolation (PODI) \cite{bui} 
which has been widely adopted for the development of ROM not only in industrial contexts \cite{demo1, demo2019complete, demo2018efficient, salmoiraghi2018free, dolci2016proper, shah2022finite} 
but also in biomedical engineering \cite{lvad, siena2022fast, pier, siena2022data}. It has been also used within a CFD-DEM framework but only for the time reconstruction \cite{shuo, shuo2, mori}. Here we consider two variants of the PODI approach: the global PODI and the local PODI. Both techniques have been employed in the Python package EZyRB \cite{demo2018ezyrb}. 

\subsection{The global PODI}

The global PODI method is based on the following \emph{offline}-\emph{online} paradigm: 

\begin{itemize}
\item during the offline phase: a set of time-dependent high-fidelity solutions is collected for a wide range of parameter values and a global reduced basis for the space of the reduced solutions is extracted
via POD. In this stage, the interpolation 
is performed in order to establish the relationship between time and parameters, and coefficients of the reduced solutions (modal
coefficients). Due to the large number of degrees of freedom, the offline phase is computationally expensive. However, it is carried out only once.
\item during the online phase: the modal coefficients for every new time and parameter instance are quickly obtained from the interpolation model. 
The reduced solution is given by the linear
combination of the reduced basis functions with the modal coefficients as weights (see eq. \eqref{eq:ROM_1}). 
\end{itemize}

Next, we are going to describe the building blocks of the above algorithm. Let $\Phi (t^i, \bm{\pi}^j)$ with $i = 1, \dots, N_t$ and $j = 1, \dots, N_k$ be the full order solutions obtained for different time instants $t^i$ and values of the parameter vector $\bm{\pi}^j$. Then we detect $N_s = N_t \cdot N_k$ input-output pairs $\{(t^i, \bm{\pi}^j), \Phi (t^i, \bm{\pi}^j)\}$ and we collect the snapshots in the matrix $\mathbb{\bm{S}} = [\Phi(t^1, \bm{\pi}^1), \dots, \Phi(t^{N_t}, \bm{\pi}^{N_k})]$. The application of the singular value decomposition to the matrix $\mathbb{\bm{S}}$ provides:\\
    \begin{equation}
    \mathbb{\bm{S}} = \U \bm{\Sigma} \bm{V}^T,
    \end{equation}
    where $\U \in \mathbb R^{{N_c}\times {N_s}}$ and $\bm{V} \in R^{{N_s}\times {N_s}}$ are the matrices composed by the left singular vectors  and right singular vectors, respectively, whilst $\bm{\Sigma} \in \mathbb R ^{{N_c}\times {N_s}}$ is the diagonal matrix containing the eigenvalues $\sigma_i$. 
    We are going to compute 
    $N_\Phi^r << N_s $ modes by minimizing the distance between the snapshots and their projection onto the space spanned by the reduced basis itself, i.e.:\\
    \begin{equation}
    argmin_{\U_{N_\Phi^r}} \| \mathbf\mathbb \mathbb{\bm{S}} - \U_{N_\Phi^r} \U_{N_\Phi^r}^T \mathbf\mathbb \mathbb{\bm{S}} \|_F \quad \text{with} \quad \U_{N_\Phi^r}^T\U_{N_\Phi^r} = \bm{I},
    \label{mode1}
    \end{equation}
where $\| \|_F$ is the Frobenius norm and $\U_{N_\Phi^r}$ is the matrix $\U$ truncated to the first $N_\Phi^r$ columns representing our POD space. 
    
Typically, the value of $N_\Phi^r$ is commonly chosen to meet a user-provided threshold $\delta$ for the cumulative
energy of the eigenvalues:
    \begin{equation}
    \frac{\sum_{i=1}^{N_\Phi^r}\sigma_i^2}{\sum_{i=1}^{N_s}\sigma_i^2} \ge \delta.
    \label{k_equation}
    \end{equation}
After constructing the POD space, we can approximate the input snapshots by using eq. \eqref{eq:ROM_1}:\\
    \begin{equation}
    \Phi(t^i, \bm{\pi}^j) \approx \sum_{L=1}^{N_\Phi^r}\chi_L (t^i, \bm{\pi}^j) \phi_L, \quad \text{with} \quad i = 1, \dots, N_t \quad \text{and} \quad j = 1, \dots, N_k, 
    \end{equation}
    where the modal coefficients $\chi_L (t^i, \bm{\pi}^j)$ are the elements of the matrix $\bm{C} = \U_{N_{\Phi}^r}^T \bm{S} \in \mathbb{R}^{N_\Phi^r \times N_s}$. Then the ROM is built using as input-output data the pairs $\{(t^i, \bm{\pi}^j), \chi_L(t^i, \bm{\pi}^j)\}$ associated to the $L$-th POD mode $\phi_L$ with $L = 1, \dots, N_\Phi^r$, i.e. we impose that 

\begin{equation}
    \mathbb{A}_L(t^i, \bm{\pi}^j) = \chi_L(t^i, \bm{\pi}^j), 
    \label{Al_2}
    \end{equation}
where 

 \begin{equation}
    \mathbb{A}_L(t^i, \bm{\pi}^j) = \sum_{m=1}^{N_t} \sum_{n=1}^{N_k} w_{L,m,n} \zeta_{L,m,n}(\| (t^i, \bm{\pi}^j) - (t^m, \bm{\pi}^n) \|). 
    \label{Al}
    \end{equation}

   
    Here $\zeta_{L,m,n}$ are the Radial Basis Functions \cite{buhmann2003radial}, 
    which we chose as Gaussian functions centered in $(t^m, \bm{\pi}^n)$ 
    and $w_{L,m,n} \in \mathbb{R}^{N_s}$ are unkown weights. Eq. \eqref{Al} can be 
     reformulated in terms of a  linear system\\
    \begin{equation}
    \bm{Z}_L \bm{w}_L = \bm{\chi}_L, 
    \label{sys}
    \end{equation}
    to be solved to obtain the weights $\bm{w}_L$ for every value of $L = 1, \dots, N_\Phi^r$ once at all in the offline phase.\\
Then, in the online phase, for any new pair time-parameter $(t^\star, \bm{\pi}^\star)$, the approximated modal coefficients $\mathbb{A}_L(t^\star, \bm{\pi}^\star)$ are calculated from eq. \eqref{Al} and the ROM solution is computed as:\\
    \begin{equation}
    \Phi_r(t^\star, \bm{\pi}^*) = \sum_{L=1}^{N_\Phi^r} \mathbb{A}_L(t^\star, \bm{\pi}^\star)\phi_L.               
    \end{equation}


\subsection{The local PODI}
Unlike the global PODI, here a POD basis is computed for each parameter in the
training set and the basis functions for new parameter values are found via RBFs interpolation of the basis functions associated to the training set. 
This method draws inspiration from \cite{pawar2020evolve} where local basis functions are used instead of global ones in a POD-Galerkin framework for CFD problems by providing substantial speed-up and computational savings as well as a greater accuracy. However, at the best of our knowledge, such approach is here used for the first time in a data-driven scenario and for multiphysics problems.  It should be noted that local PODI and global PODI coincide when only time reconstruction is considered. 

The local PODI method is based on the following \emph{offline}-\emph{online} paradigm: 

\begin{itemize}
\item during the offline phase: a set of time-dependent high-fidelity solutions is collected for a wide range of
parameter values and a local basis for the space of the reduced solutions associated to each parameter in the training set is extracted
via POD. In this stage, the interpolation 
is performed in order to establish the relationship between time and parameters, and modal coefficients of the reduced solutions. Moreover, an interpolation of the basis functions associated to the training set is also carried out. Due to the large number of degrees of freedom, the offline phase is computationally expensive. However, it is carried out only once.
\item during the online phase: the modal coefficients for every new time and parameter instance as well as the basis functions are quickly obtained from the interpolation models. 
The reduced solution is given by the linear
combination of the reduced basis functions with the modal coefficients as weights (see eq. \eqref{eq:ROM_1}). 
\end{itemize}

Then the key differences with respect to the global PODI are the following:

\begin{itemize}
\item instead to work with a single matrix $\bm{S}$ containing all the $N_s = N_t \cdot N_k$ snapshots, we collect $N_k$ matrices, one for each training value of the parameter $\bm{\pi}^j$: $\mathbb{\bm{S}}^j = [\Phi(t^1, \bm{\pi}^j), \dots,  \Phi(t^{N_t}, \bm{\pi}^{j})]$. Then the singular value decomposition is applied to each matrix $\mathbb{\bm{S}}^j$.

\item The POD spaces computed in this way are then interpolated in order to compute the basis functions for any new parameter $\bm{\pi}^\star$. 
\end{itemize}

\section{Numerical results}\label{sec:res}
\graphicspath{{./img/}}

This section presents several numerical results in order to test the performance of our ROM approach. 

\begin{figure} 
    \centering  \includegraphics[width=50mm,scale=0.2]{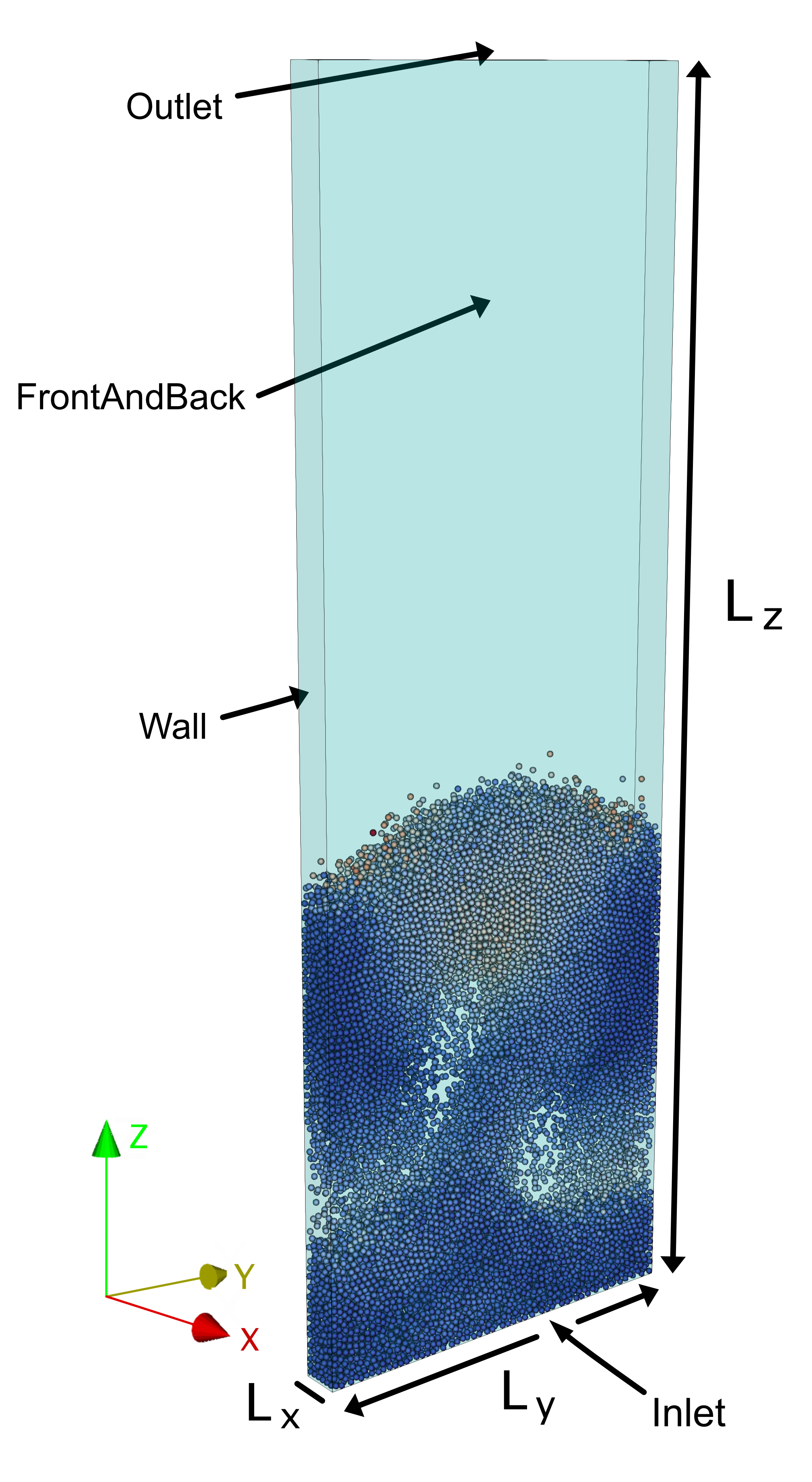}
        \caption{
Sketch of the computational domain. The particles flow refers to $t = 1$ s and is colored by the velocity magnitude. The labels identify the surfaces of the domain on which different boundary conditions are employed. 
}
\label{fig:BC}
\end{figure}

As the benchmark case, we consider a system of gas-solid fluidized bed where the small particles are transported by the carrier gas flow \cite{goldschmidt2001hydrodynamic, fernandes}. The computational domain consists of a rectangle with dimensions $L_x \times L_y \times L_z = 15 \times 150 \times 450$ $mm$. It is discretized using $N_x \times N_y \times N_z = 2 \times 30 \times 90$ cells. See \fig{fig:BC} for a sketch of the geometry.
The gas dynamics is solved only in a 2D framework along $y$ and $z$ directions whilst the particle motion is fully 3D \cite{goldschmidt2001hydrodynamic}. 

For the fluid phase, we set  $\mu_f = 1e-05 \: Pa \cdot s$ and $\rho_f = 1000  \: Kg/m^3$.
Concerning the boundary conditions, we refer to the labels reported in \fig{fig:BC} indicating the boundaries of our computational domain. We enforce no-slip boundary conditions on \emph{Wall} faces. On the other hand, the \emph{FrontAndBack} faces are set to symmetry. At the \emph{Outlet} boundary, a mixed condition is considered employing an homogeneous Neumann boundary condition when there is outflow, otherwise a null normal velocity, in order to prevent backflow. 
Finally, a non homogenenous Dirichlet boundary condition is employed at the \emph{Inlet} face where the actual interstitial velocity (i.e. the upward velocity of the fluid flow through the open area between the particles) is calculated by dividing the specified velocity value ($1.875 \: m/s$)  by the fluid volume fraction. 

At the beginning of simulations $n_p = 24750$ spherical particles with $d_p = 0.0025\: m$ and $\rho_p = 2488.32 \: kg / m^3$ are distributed uniformly in the domain with an initial bed height of $300 \: mm$. Then we have $Stk = 300$ (see eq. \eqref{Stk}). We start the simulations from fluid and particles at rest. 
We let the system evolve until time $T =5$ s using a fixed time-step $\Delta t = 2e-5$ s which is chosen based on maximum Courant number to keep it always below the unity for sake of numerical stability. 

\begin{figure}\vspace{2cm}
    \centering
    \begin{subfigure}{0.75\textwidth}
        \includegraphics[width=\linewidth]{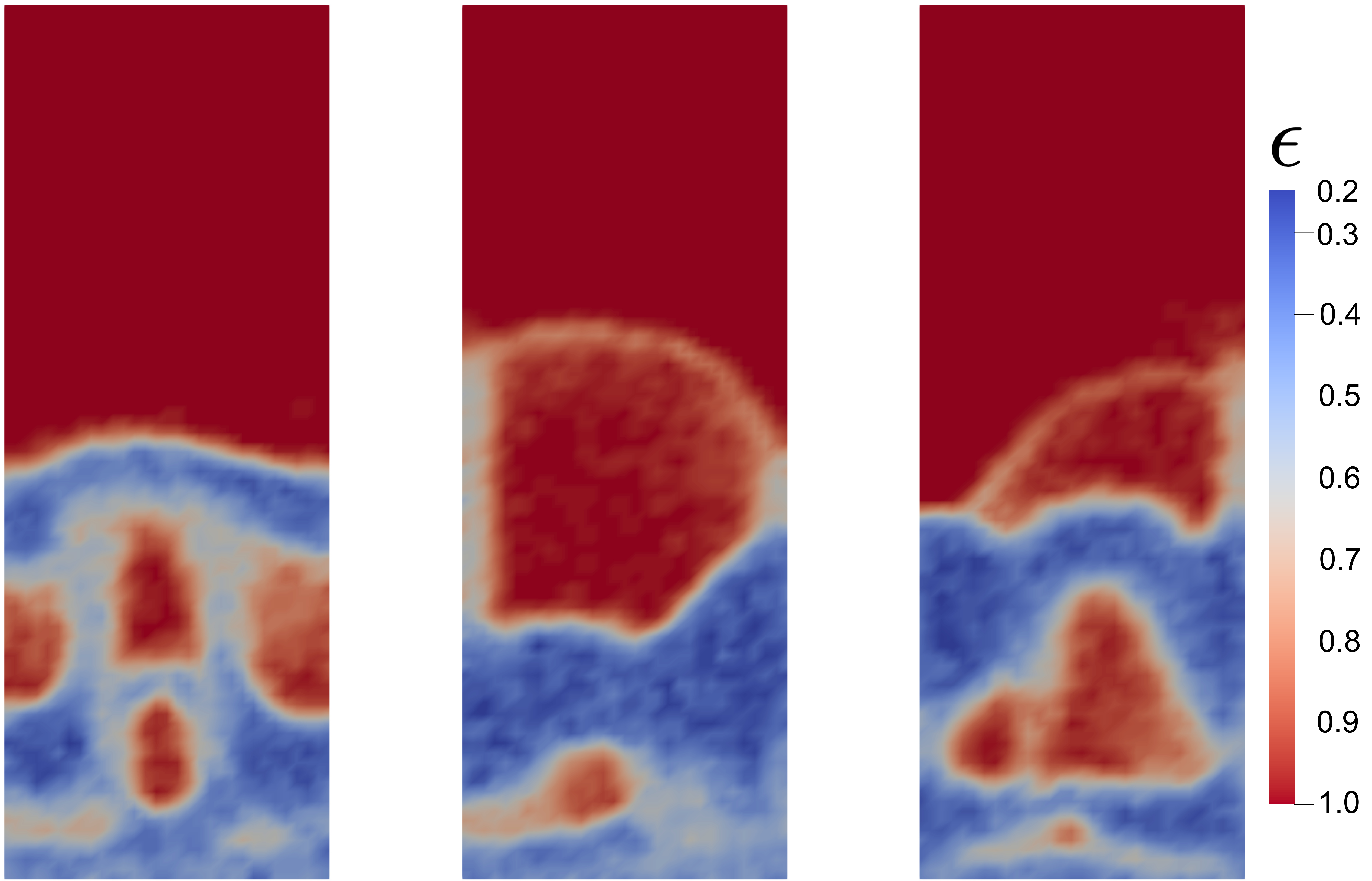}
        \subcaption{Time evolution of the fluid volume fraction: $t = 1$ s (left), $t = 2.5$ s (center) and $t = 4$ s (right).}
        
        \label{fig:Eurlerian_FOMSol}
    \end{subfigure}

\medskip
       \begin{subfigure}{0.8\textwidth}
        \includegraphics[width=\linewidth]{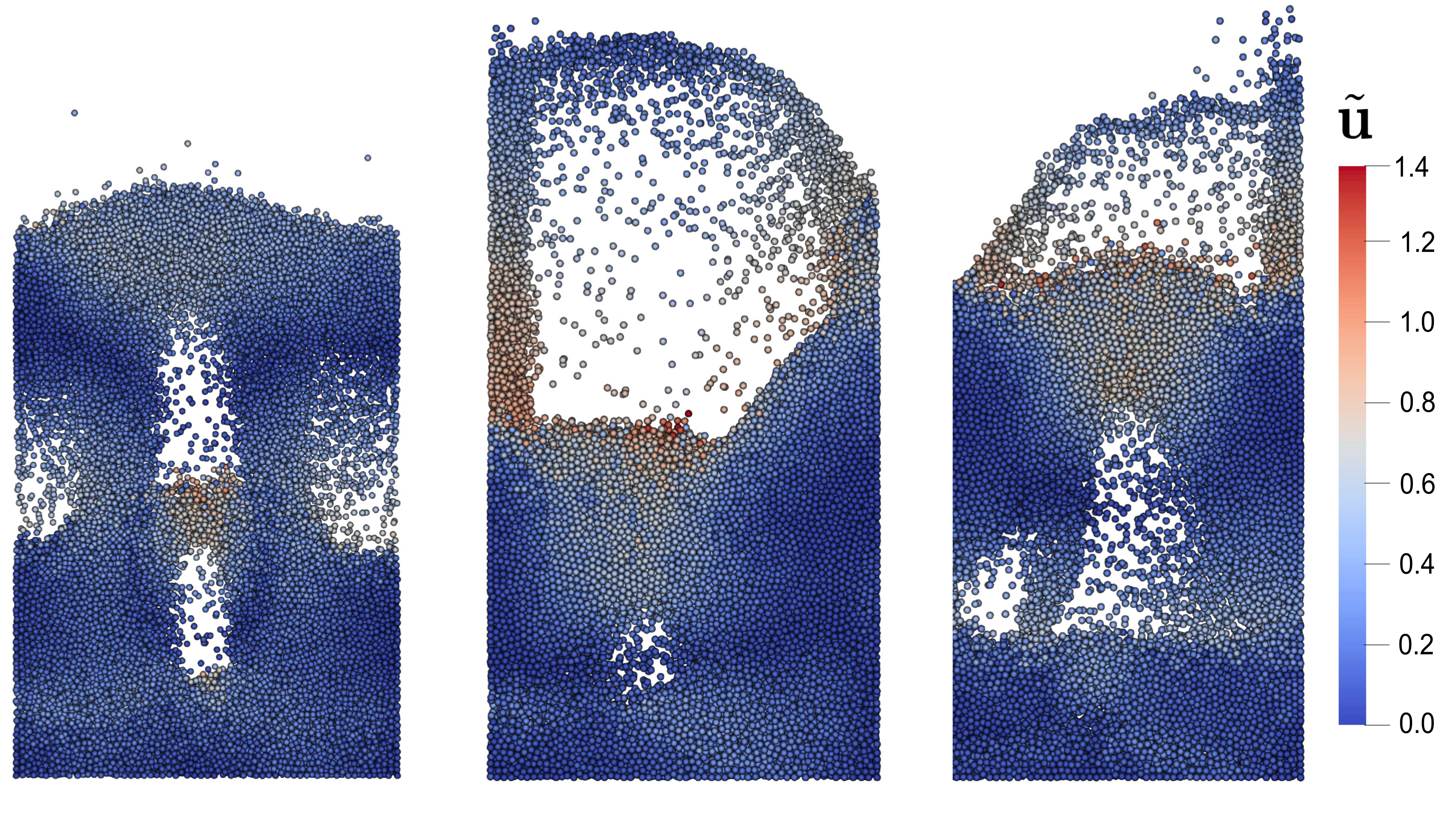}
        \subcaption{Time evolution of the particles distribution: $t = 1$ s (left), $t = 2.5$ s (center) and $t = 4$ s (right).}
        
        \label{fig:Lagrang_FOMSol}
    \end{subfigure}
    \caption{FOM solution for the fluid volume fraction $\epsilon$ (a) and particle position (b) at times $t = 1$ s (first column), $t = 2.5$ s (second column) and $t = 4$ s (third column). The particles are colored based on the magnitude of their velocity.}
\label{fig:full_order}
\end{figure}
Qualitative illustrations of instantaneous flow field are depicted in \fig{fig:full_order}  where the first row shows the distribution of the fluid volume fraction whilst the second one the position of the particles for three different time instances: $t = 1$ s, $t = 2.5$ s and $t = 4$ s from left to right. 
Initially all the particles are evenly located; then, when the gas velocity reaches the fluidization velocity, all the particles are suspended by the upward gas \cite{goldschmidt2001hydrodynamic, fernandes}.

As we see, we are dealing with a very complex two-phase system 
in which we expect that a large number of POD modes are to be retained to capture and predict properly its dynamics. Firstly, we are going to investigate the performance of the ROM model in the reconstruction
of the time evolution of the flow field in Sec. \ref{sec:time_rec}. Then parametrization of the
Stokes number is introduced in Sec. \ref{sec:Stokes_param}.

\subsection{Time reconstruction}\label{sec:time_rec}
\hfill \break

The starting high-fidelity database for all the numerical tests presented in this section consists of 500 FOM snapshots which are collected every 0.02 s.

\subsubsection{Reconstruction of the Eulerian field}
\begin{figure}[ht]
    \centering
        \includegraphics[width=80mm,scale=0.5]{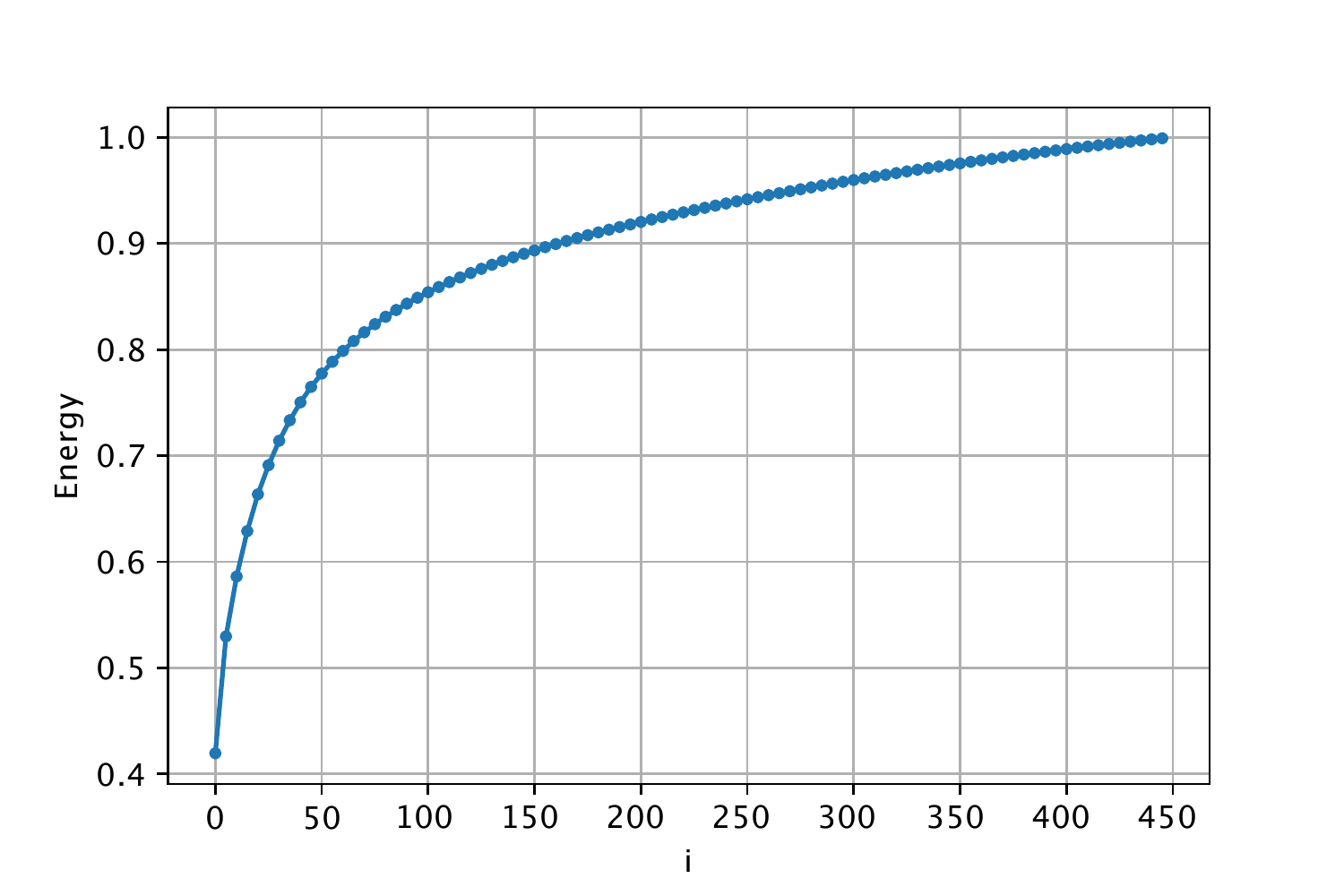}
        \caption{ROM validation - time reconstruction of the Eulerian field: cumulative eigenvalues for the fluid volume fraction $\epsilon$.}

        \label{fig:modes_energy_alpha}
\end{figure}

\hfill \break
In the first numerical experiment, we train our ROM model by using the 90$\%$ of the initial database (i.e., 450 snapshots) whilst the remaining 10\% (i.e., 50 snapshots) is used for the validation. We highlight that the FOM snapshots belonging to the training and validation set are randomly selected. 
In \fig{fig:modes_energy_alpha} it is shown the plot of the cumulative eigenvalues for the fluid volume fration $\epsilon$. 
The curve shows a very slow decay: at least 150 POD modes are needed to capture 90$\%$ of the system's energy. 


\begin{figure}
    \centering
    \begin{subfigure}{0.65\textwidth}
        \includegraphics[width=\linewidth]{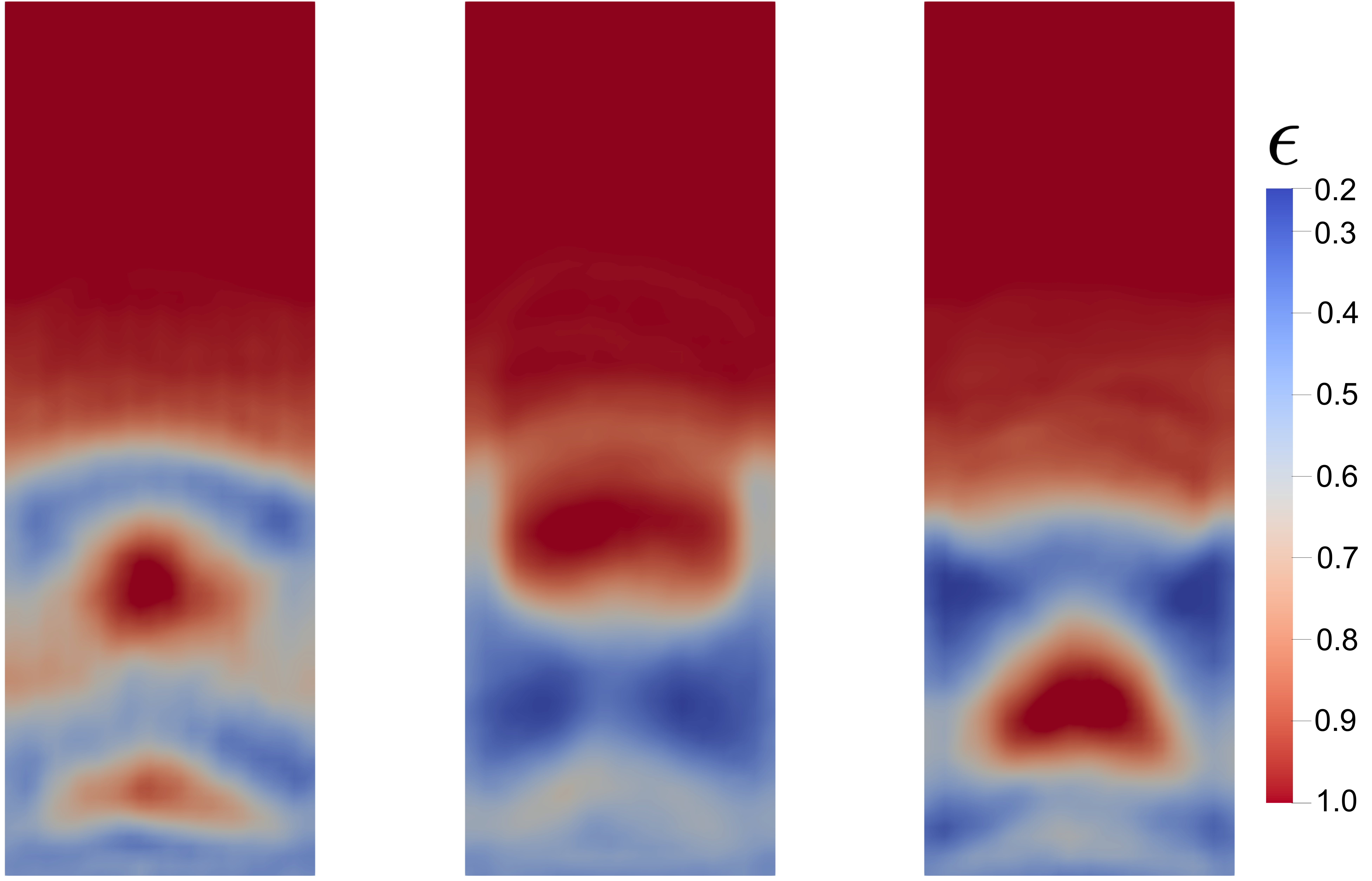}
        \subcaption{Time evolution for $\delta = 50\%$ (4 POD modes): $t = 1$ s (left), $t = 2.5$ s (center) and $t = 4$ s (right).}
        \label{fig:RecSol0.5}
    \end{subfigure}
\medskip
    \begin{subfigure}{0.65\textwidth}
        \includegraphics[width=\linewidth]{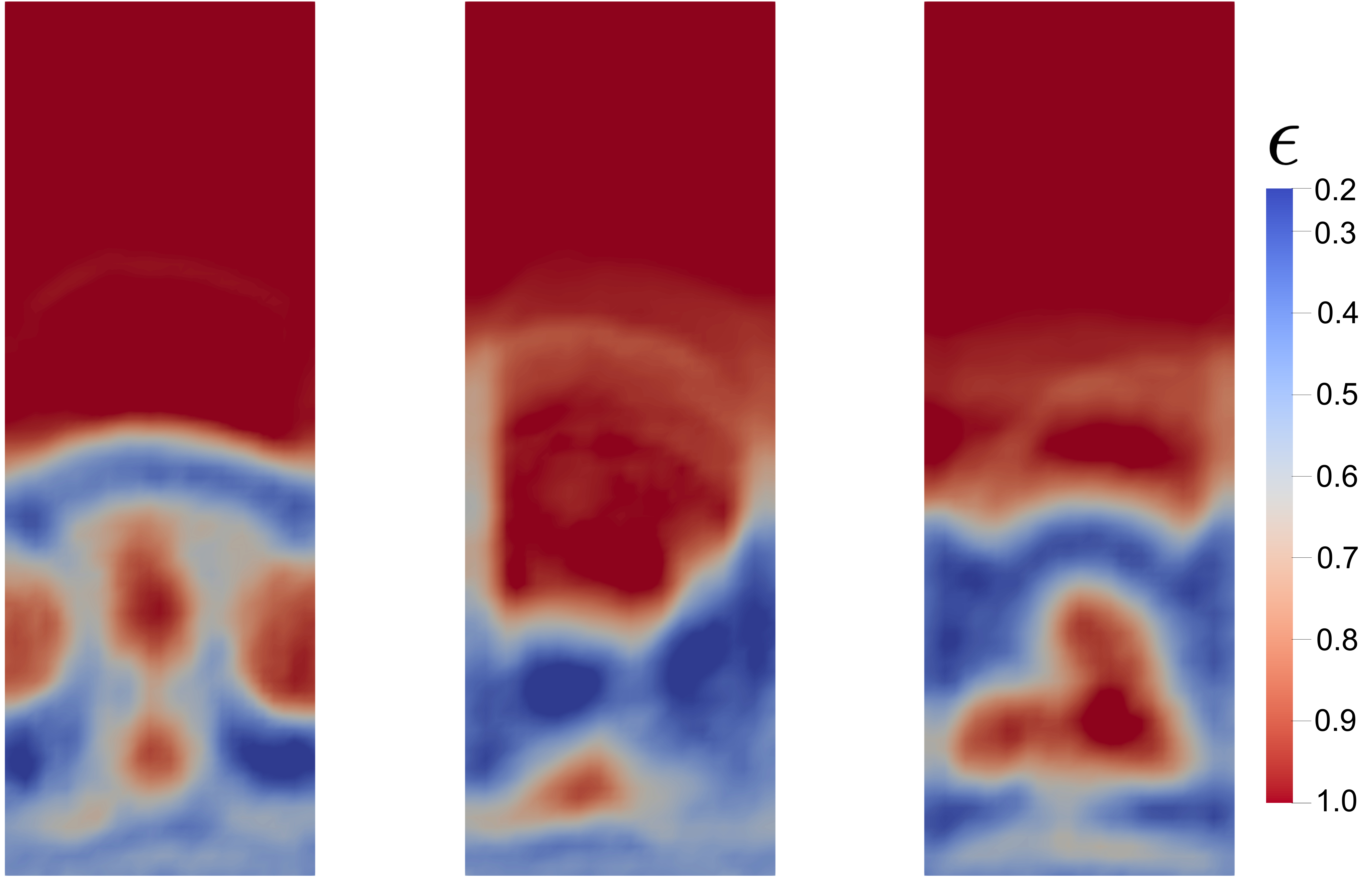}
        \subcaption{Time evolution for $\delta = 70\%$ (60 POD modes):  $t = 1$ s (left), $t = 2.5$ s (center) and $t = 4$ s (right).}
        \label{fig:RecSol0.7}
    \end{subfigure}
    \medskip
    \begin{subfigure}{0.65\textwidth}
        \includegraphics[width=\linewidth]{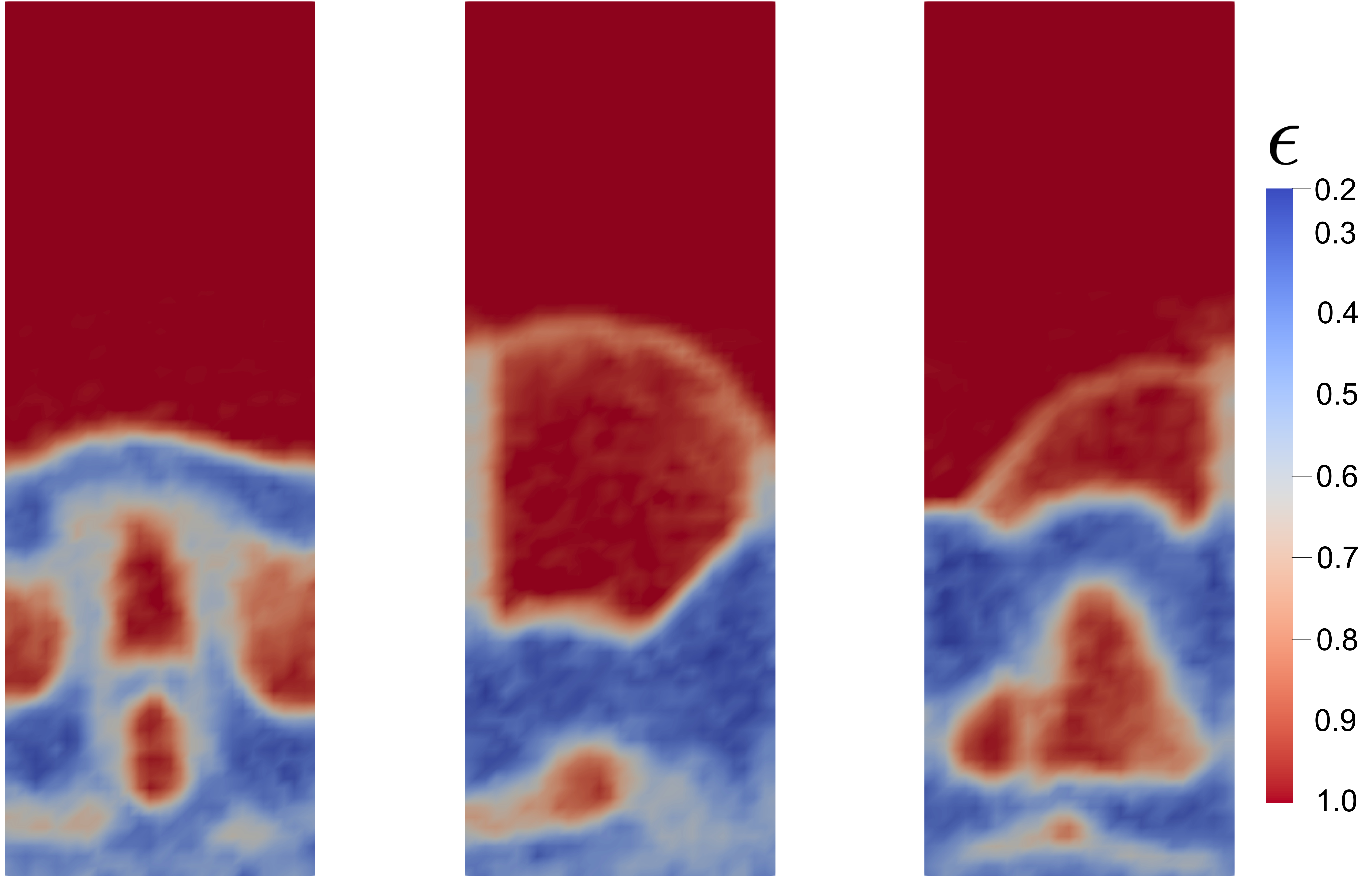}
        \subcaption{Time evolution for $\delta = 90\%$ (150 POD modes):  $t = 1$ s (left), $t = 2.5$ s (center) and $t = 4$ s (right).}
        \label{fig:RecSol0.9}
    \end{subfigure}    
\caption{ROM validation - time reconstruction of the Eulerian field: fluid volume fraction $\epsilon$ computed by ROM for three different energy thresholds $\delta = 50\%$ (a), $\delta = 70 \%$ (b) and $\delta = 90 \%$ (c) at times $t = 1$ s (first column), $t = 2.5$ s (second column) and $t = 5$ s (third column).} 
\label{fig:Vis_VoidFraction}
\end{figure}
Then we provide a qualitative comparison between FOM (\fig{fig:Eurlerian_FOMSol}) and ROM (\fig{fig:Vis_VoidFraction}). 
We consider three different energy thresholds: $\delta = 50\% $, $\delta = 70\% $ and $\delta = 90\% $  corresponding to 4, 60 and 150 POD modes, respectively. 
At first look, 
we see that, even by considering only the 50$\%$ of the energy (\fig{fig:RecSol0.5}), the ROM is able to reconstruct the main patterns of the flow field, although some structures are not detected. 
Of course more accurate results could be obtained with a larger amount of energy/number of POD modes: see ROM solutions obtained by retaining the 70$\%$ (\fig{fig:RecSol0.7}) and 90\% (\fig{fig:RecSol0.9}) of the energy. 

Now we consider another numerical experiment in order to provide a more quantitative comparison. We are going to perform a sensitivity analysis with respect to the cardinality of the training set, i.e. the number of high-fidelity snapshots that we take into account to train our ROM, and for each training set we refer to different energy levels. We are interest in the time evolution of the $L^2$-norm relative error between FOM and ROM:
\begin{equation}\label{eq:error1}
E_{\epsilon}(t) = 100 \cdot \dfrac{||\epsilon_h(t) - \epsilon_r(t)||_{L^2(\Omega)}}{||{\epsilon_h}(t)||_{L^2(\Omega)}},
\end{equation}
where $\epsilon_h$ is the fluid volume fraction computed with the FOM and $\epsilon_r$ is the corresponding field computed with the ROM.
Three different configurations are considered: in the first case, the 50$\%$ of the high-fidelity solutions contained in the initial database is used to train the ROM model, then the 70$\%$ and finally the 90$\%$. 
\begin{figure}[ht!]\centering
\subfloat[50 $\%$ of FOM database used for ROM training]{\label{fig:50PercentTraining}\includegraphics[width=.5\linewidth]{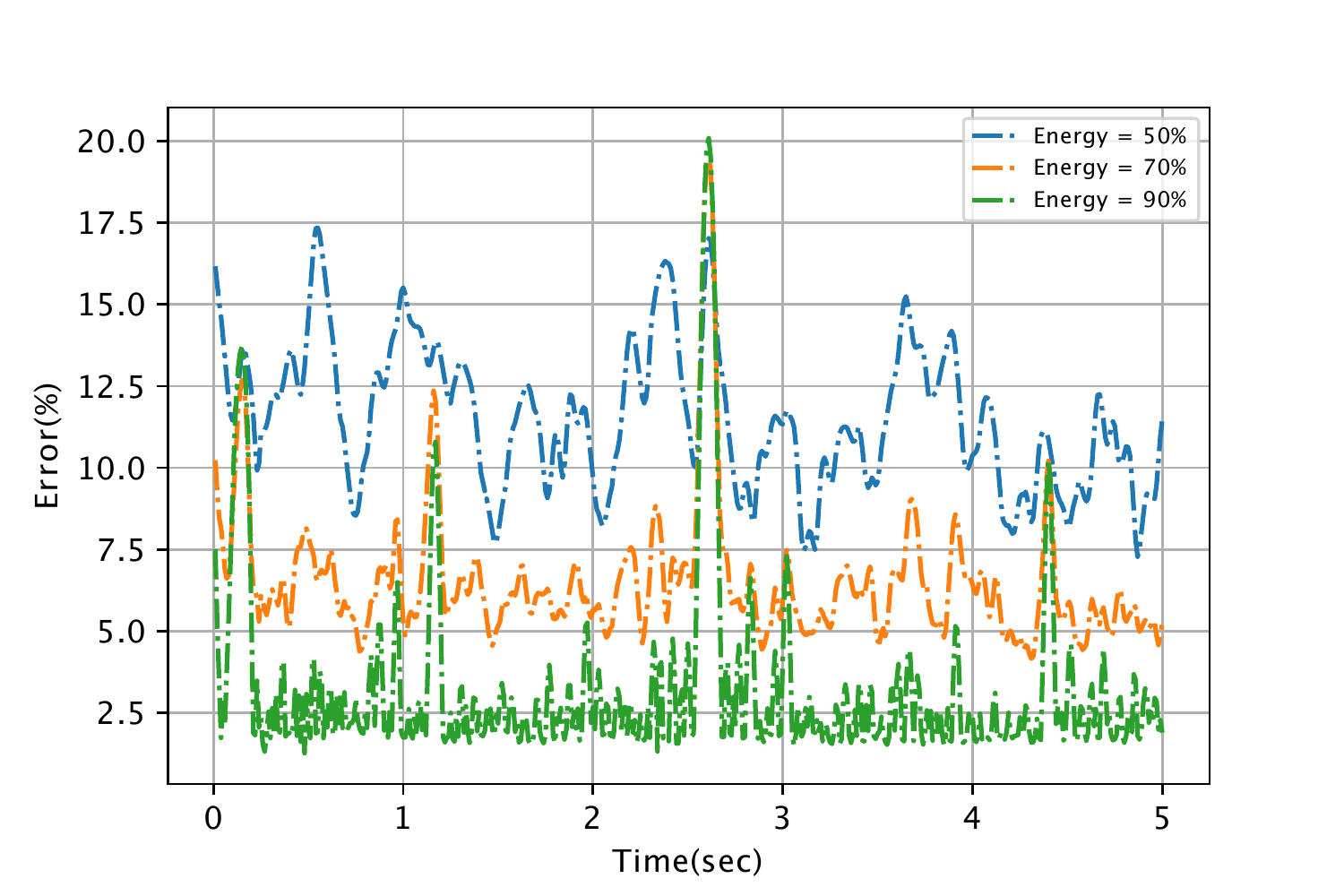}}\hfill
\subfloat[70 $\%$ of FOM database used for ROM training]{\label{fig:70PercentTraining}\includegraphics[width=.5\linewidth]{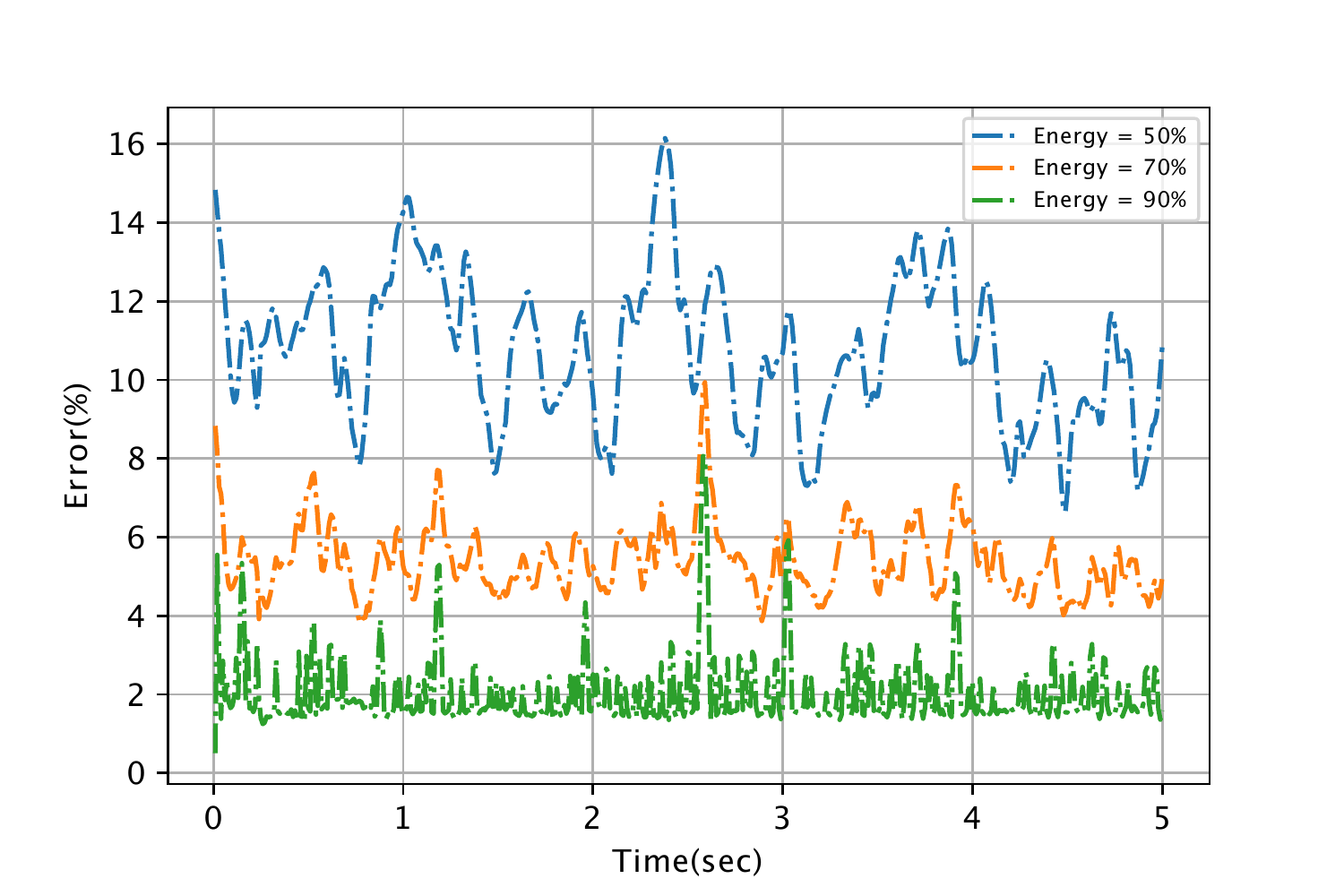}}\par 
\subfloat[90 $\%$ of FOM database used for ROM training]{\label{fig:90PercentTraining}\includegraphics[width=.5\linewidth]{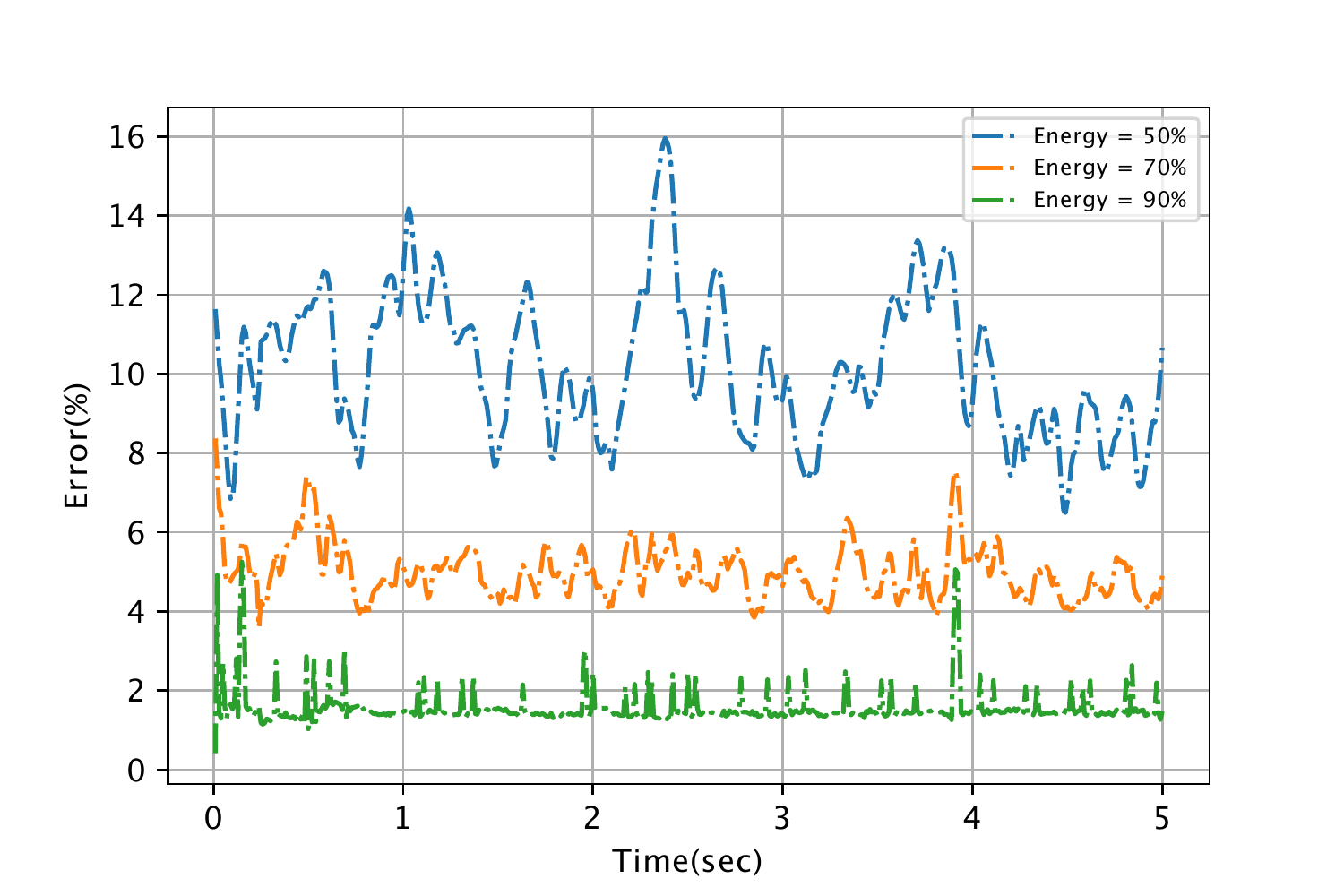}}
\caption{ROM validation - time reconstruction of the Eulerian field: time evolution of the $L^2$-norm relative error for different training set and energy levels.
}
\label{fig:Error_in_time}
\end{figure}
In all the cases, three different values of $\delta$ are set: 50\%, 70\% and 90\%. 
In \fig{fig:Error_in_time} we observe that as expected, at increasing of the training set size, 
the error reaches lower values and some scattered oscillations are damped, especially when the 90\% of the energy is retained. 
On the other hand, at a given training set, the highest energy threshold corresponds to the lowest error.
This result is in agreement with what was already qualitatively observed in \fig{fig:Vis_VoidFraction}. 
For further comparison, some statistics of the $L^2$-norm error, its maximum and minimum value as well as its time average value for the richer training set, are reported in  
\tab{tab:VoidFrac_error}. 
\begin{table}[h!]
\centering
 \begin{tabular}{||c |c |c |c||} 
 \hline
Energy threshold  & $\delta=50 \%$ & $\delta=70\%$ & $\delta=90\%$ \\ [0.5ex] 
 \hline\hline
 Mean Error (\%) & 10.2 & 5 & 1.1 \\ \hline
 Max Error (\%) & 15.9 & 8.4 & 5.38 \\ \hline
 Min Error (\%) & 6.5 & 2.6 & 0.2 \\ 
 \hline
 \end{tabular}
 \caption{ROM validation - time reconstruction of the Eulerian field: mean, maximum and minimum values of the $L^2$-norm relative error (see eq. \eqref{eq:error2}) by using the $90\%$ of the starting FOM database to train our ROM model for three different energy thresholds.} 
 \label{tab:VoidFrac_error}
\end{table}
We observe that the mean and the minimum errors decrease of about one order of magnitude when one moves from $\delta = 50\%$ to $\delta = 90\%$. The maximum error reduces about three times. Anyway, we can see that already with $\delta = 50\%$ of the energy the mean error does not overcome 10\%, demonstrating a good accuracy of the ROM. 

\subsubsection{Reconstruction of the Lagrangian field}
\hfill



Based on what obtained for the Eulerian phase by the sensitivity analysis with respect to the cardinality of the training set, for all the numerical tests presented in this subsection we train our ROM model by using the 90\% of the FOM database whilst the remaining 10\% is adopted for the validation.

\begin{figure}[ht]
\centering
\subfloat[Particle position $\bm{\widetilde{x}}$.]{\label{fig:CumEnergy_xp}\includegraphics[width=.52\linewidth]{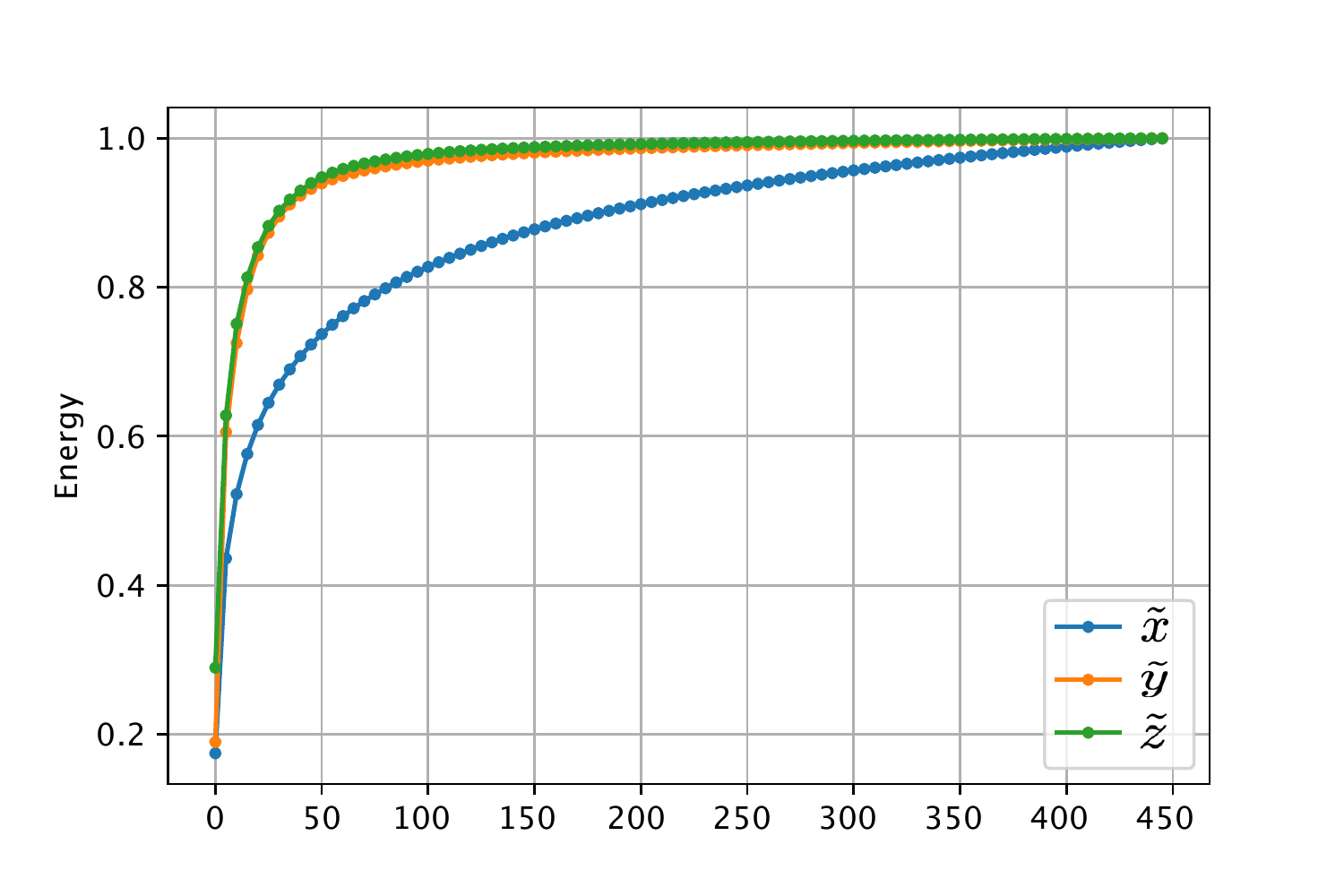}}
\subfloat[Particle velocity $\bm{\widetilde{u}}$.]{\label{fig:CumEnergy_up}\includegraphics[width=.52\linewidth]{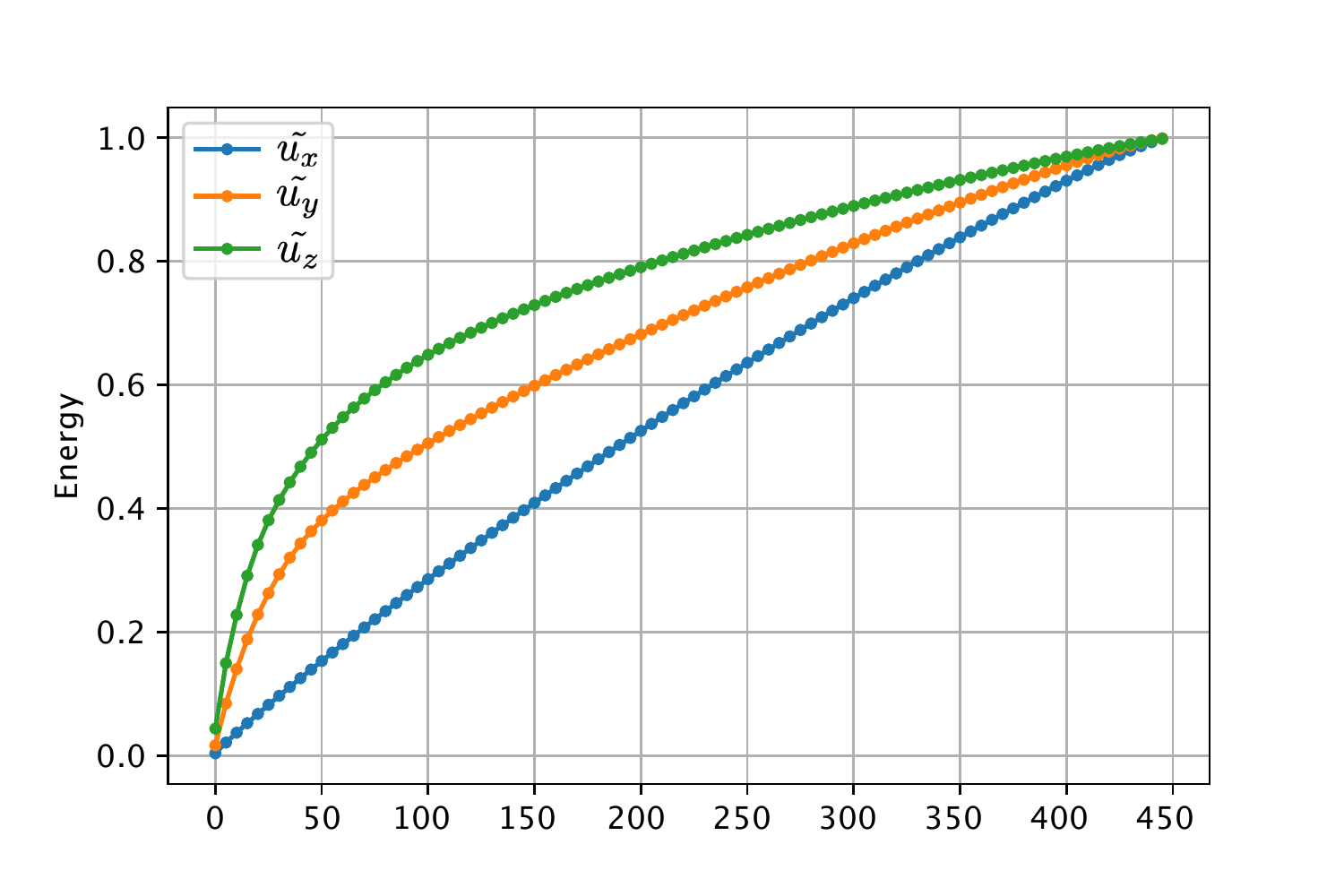}}
\caption{ROM validation - time reconstruction of the Lagragian field: cumulative eigenvalues for particle position $\bm{\widetilde{x}}$ and velocity $\bm{\widetilde{u}}$ in panel (A) and (B), respectively.} .
\label{fig:modes_time_Lag}
\end{figure}

In \fig{fig:modes_time_Lag} it is shown the plot of the cumulative eigenvalues for particle position and velocity. We can observe that $\widetilde{y}$ and $\widetilde{z}$ exhibit a faster decay compared to $\widetilde{x}$: in fact, in order to capture the 90\% of the energy, 50 POD modes are enough for the former ones whilst at least 300 POD modes are required for the latter one. The situation in even worst in terms of velocity; in fact,   $\widetilde{u_x}$ exhibits an almost linear relationship between eigenvalues and modes numbers, resulting in 400 POD modes to retain the 90$\%$ of the energy. 
Indeed, the smallness of the ratios $L_x/L_y$ and $L_x/L_z$ as well as the very low number of particles which can be located along $L_x$ ($d_p$ $\approx$ $L_x$)  
significantly limit the dynamics in the $x$ direction leading to a small scale motion which is very challenging to be recovered by ROM. Moreover, the faster convergence of the position with respect to the velocity (that is exhibited by all the three components)  could be due to the fact that the particles are very close to each other, so they collide very frequently by exchanging their momentum.
\begin{figure}
    \centering
    \begin{subfigure}{0.8\textwidth}
        \includegraphics[width=\linewidth]{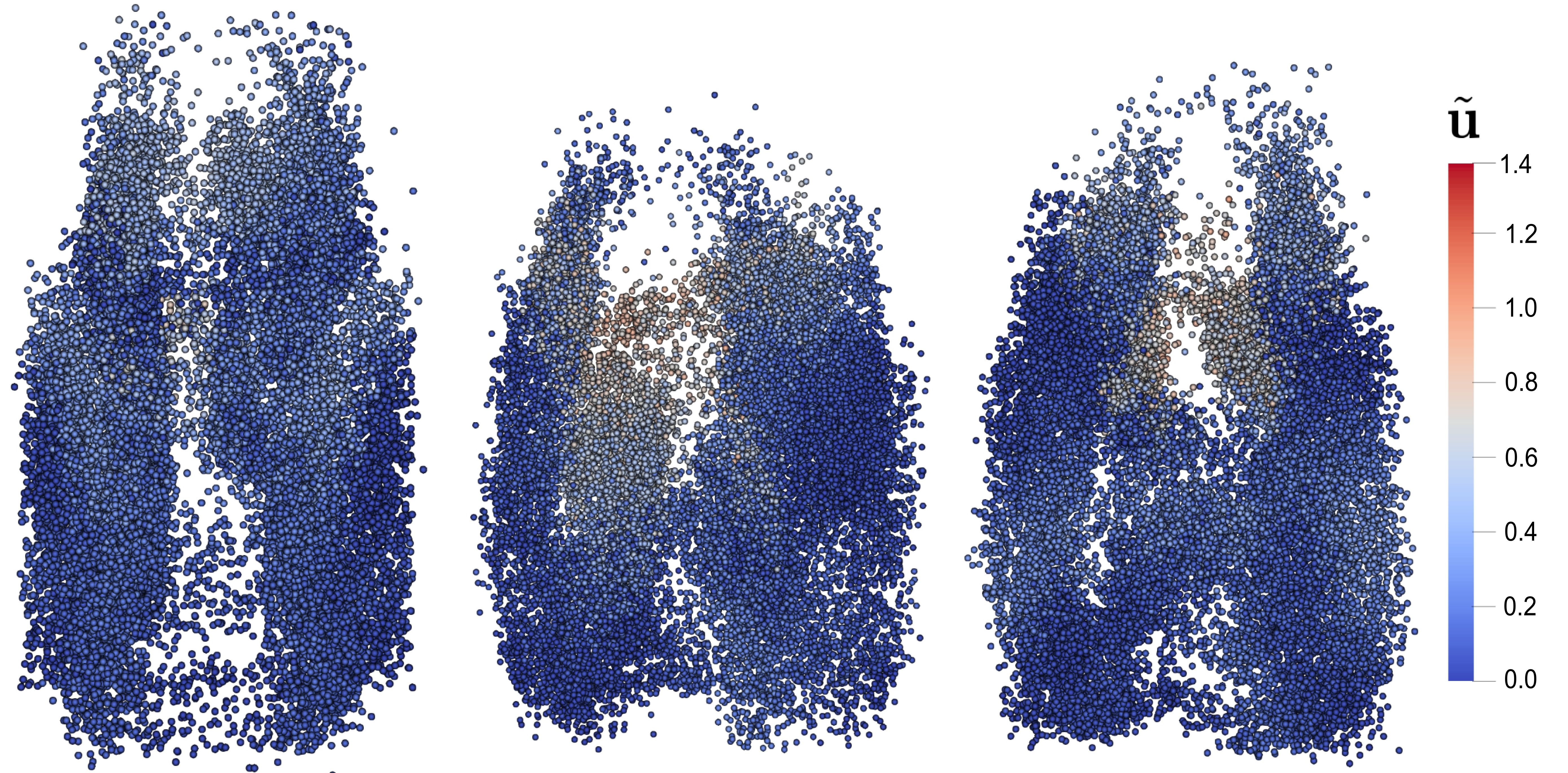}
        \subcaption{Time evolution for $\delta$ = 50\%: $t = 1$ s (left), $t = 2.5$ s (center) and $t = 4$ s (right).}
        \label{fig:RecSol0.5_Lag}
    \end{subfigure}
\medskip
    \begin{subfigure}{0.8\textwidth}
        \includegraphics[width=\linewidth]{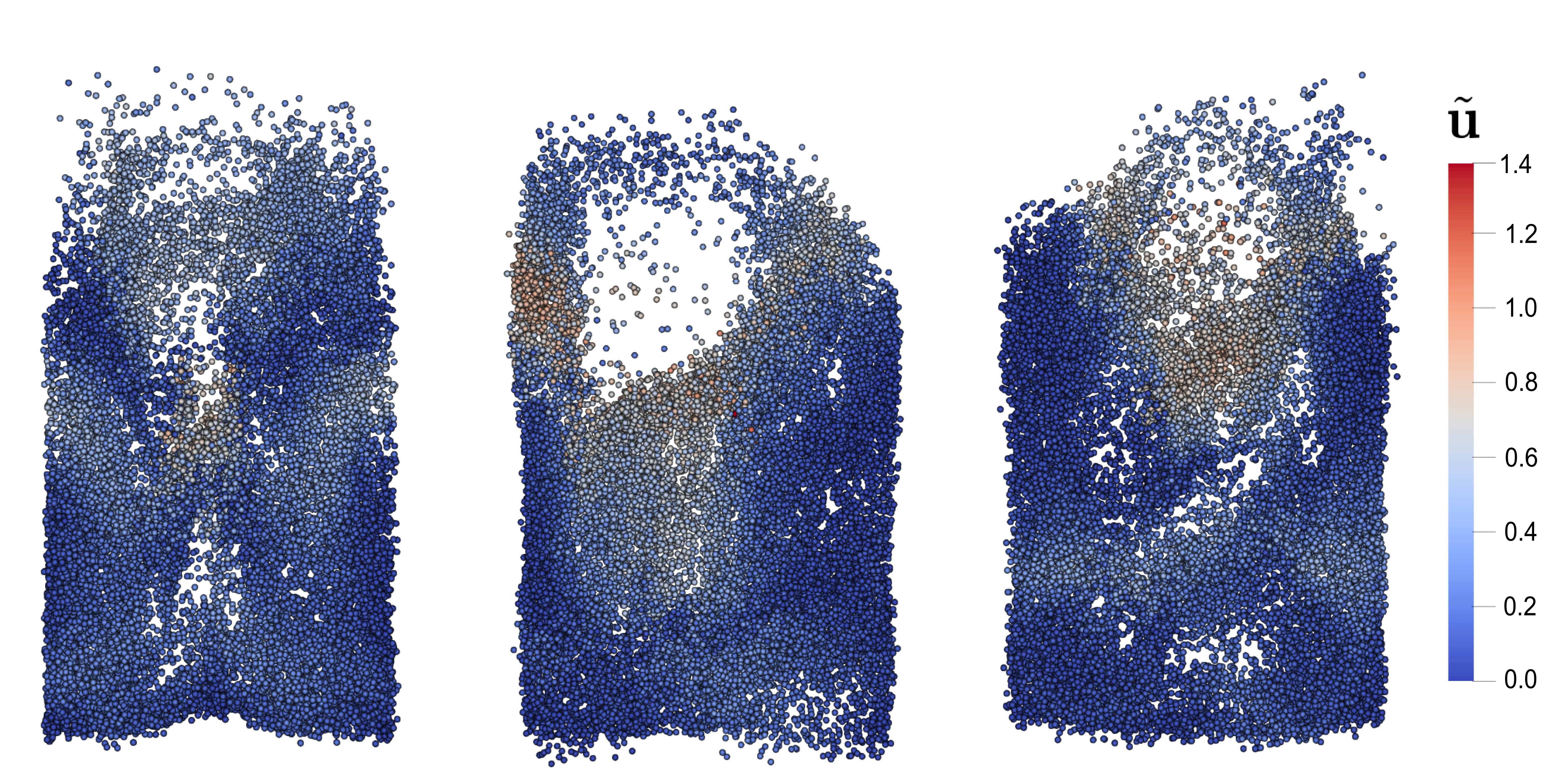}
        \subcaption{Time evolution for $\delta$ = 70\%: $t = 1$ s (left), $t = 2.5$ s (center) and $t = 4$ s (right).}
        \label{fig:RecSol0.7_Lag}
    \end{subfigure}
    \medskip
    \begin{subfigure}{0.8\textwidth}
        \includegraphics[width=\linewidth]{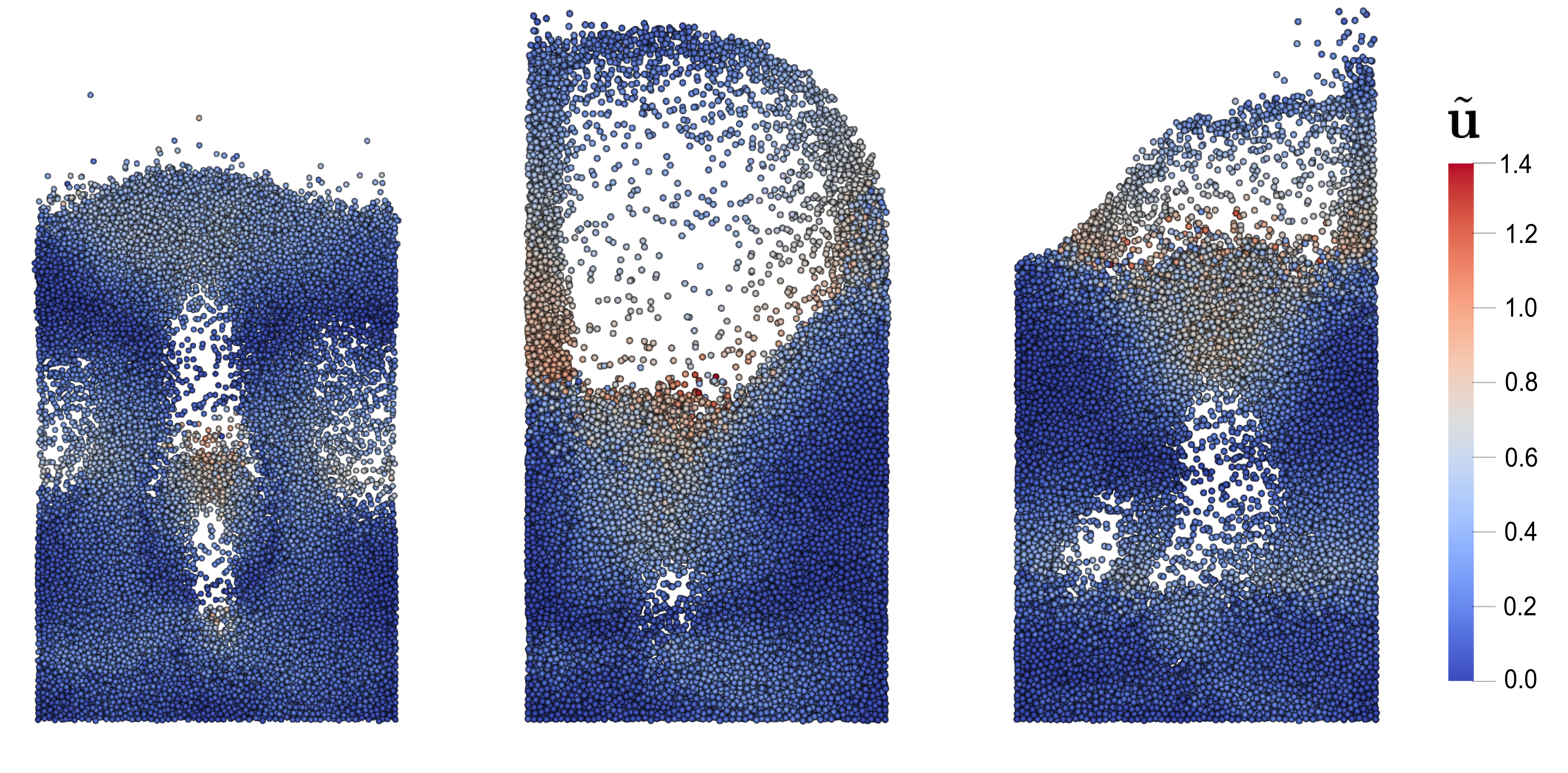}
        \subcaption{Time evolution for $\delta$ = 90\%: $t = 1$ s (left), $t = 2.5$ s (center) and $t = 4$ s (right).}
        \label{fig:RecSol0.9_Lag}
    \end{subfigure}    
    \caption{ROM validation - time reconstruction of the Lagrangian field: particle position computed by ROM for three different energy tresholds $\delta = 50\%$ (a), $\delta = 70\%$ (b) and $\delta = 90\%$ (c) at times $t = 1$ s (first column), $t = 2.5$ s (second column) and $t = 4$ s (third column). The particles are colored based on the magnitude of their
velocity.}
    \label{fig:RecSol_Lag}
\end{figure}

 
\fig{fig:RecSol_Lag} shows an illustrative representation of the time evolution of the ROM solution for particle position and velocity. 
Each Lagrangian point in \fig{fig:RecSol_Lag} represents the reconstructed solution of the three components of particle position. The color of each point depicts the magnitude of the corresponding velocity. 
As for the Eulerian case, three different energy thresholds are considered: $\delta = 50 \%$, $\delta = 70 \%$ and $\delta = 90 \%$. The corresponding POD modes numbers are reported in \tab{tab:ModeNumber_Lagrangian}.
From the comparison with \fig{fig:Lagrang_FOMSol}, we observe that for $\delta = 50\%$ the ROM is not able to detect the right trend. Furthermore, some particles are located outside the computational domain along the $y$ direction. However, the higher the energy retained/POD modes considered, the higher the accuracy of the ROM as can be observed in   \fig{fig:RecSol0.7_Lag} ($\delta = 70 \%$) and \fig{fig:RecSol0.9_Lag}  ($\delta = 90 \%$). In short, a very good accuracy of ROM to identify the discrete phase and capture its dynamic can be appreciated when $\delta = 90\%$ (\fig{fig:RecSol0.9_Lag}). 


\begin{table}[ht!]
\centering
\begin{tabular}{||c|c|c|c|c|c|c||}
\hline 
  &  \multicolumn{3}{c|}{Particle position}  &  \multicolumn{3}{c||}{Particle velocity} \\
\cline{2-7}\cline{2-7}
            &  $\Tilde{x}$ & $\Tilde{y}$ & $\Tilde{z}$& $\Tilde{u_x}$ & $\Tilde{u_y}$ & $\Tilde{u_z}$ \\ \hline
$\delta = 50 \%$  & 10  & 4  & 4  & 190 & 99  & 49   \\ \hline
$\delta = 70 \%$  & 39  & 10 & 9  & 282 & 213 & 131  \\\hline
$\delta = 90 \%$  & 183 & 33 & 31 & 384 & 355 & 313  \\\hline
\end{tabular}
 \caption{ROM validation - time reconstruction of the Lagrangian field: numbers of POD numbers associated to the energy thresholds $\delta = 50 \%$, $\delta = 70 \%$ and $\delta = 90 \%$ for each component of the particle position and velocity.}
 \label{tab:ModeNumber_Lagrangian}
\end{table}

\begin{figure}\centering
\subfloat[Particle position $\widetilde{x}$]{\label{fig:Error-diffEng_xp}\includegraphics[width=.5\linewidth]{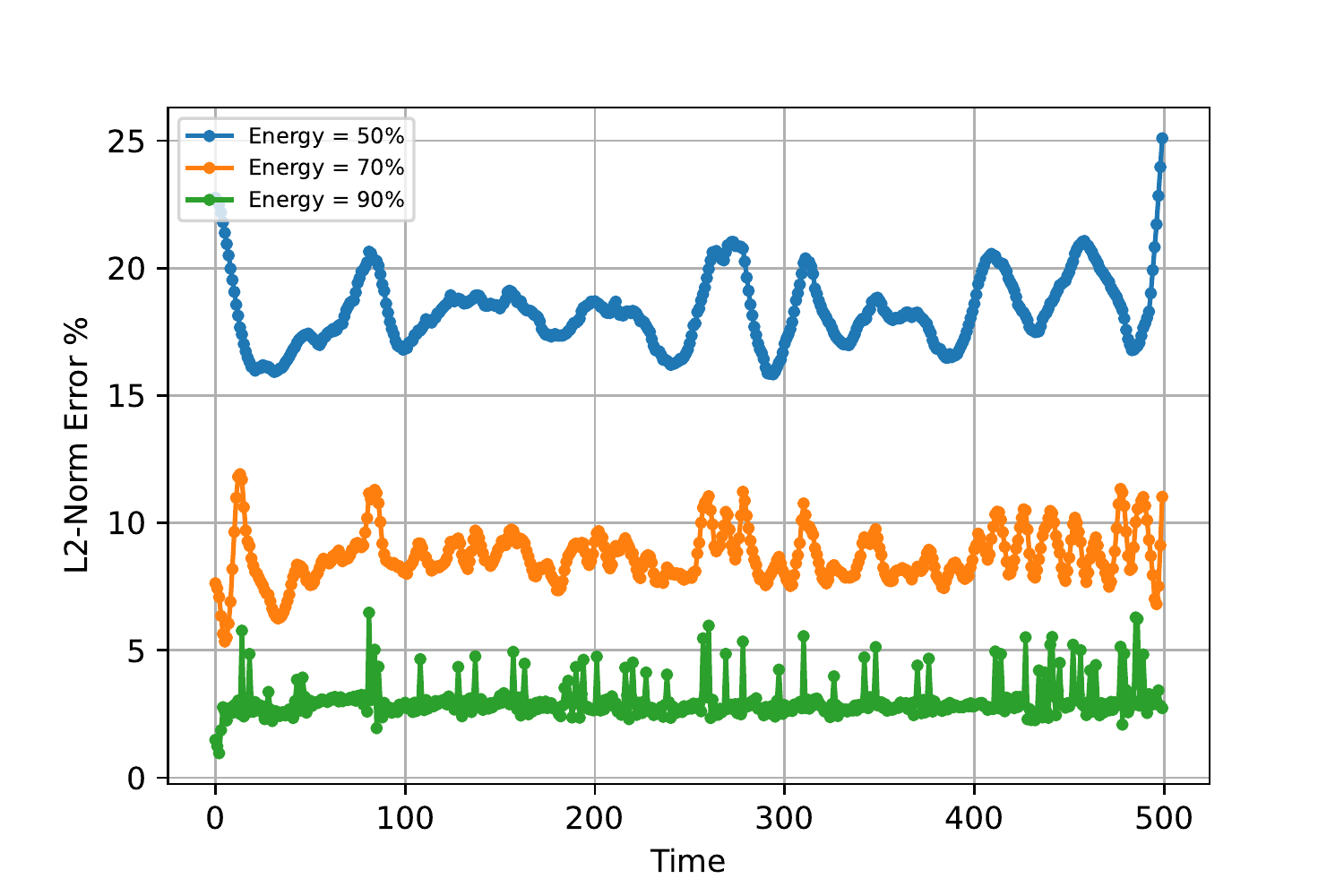}}\hfil
\subfloat[Particle position in $\widetilde{y}$]{\label{fig:Error-diffEng_yp}\includegraphics[width=.5\linewidth]{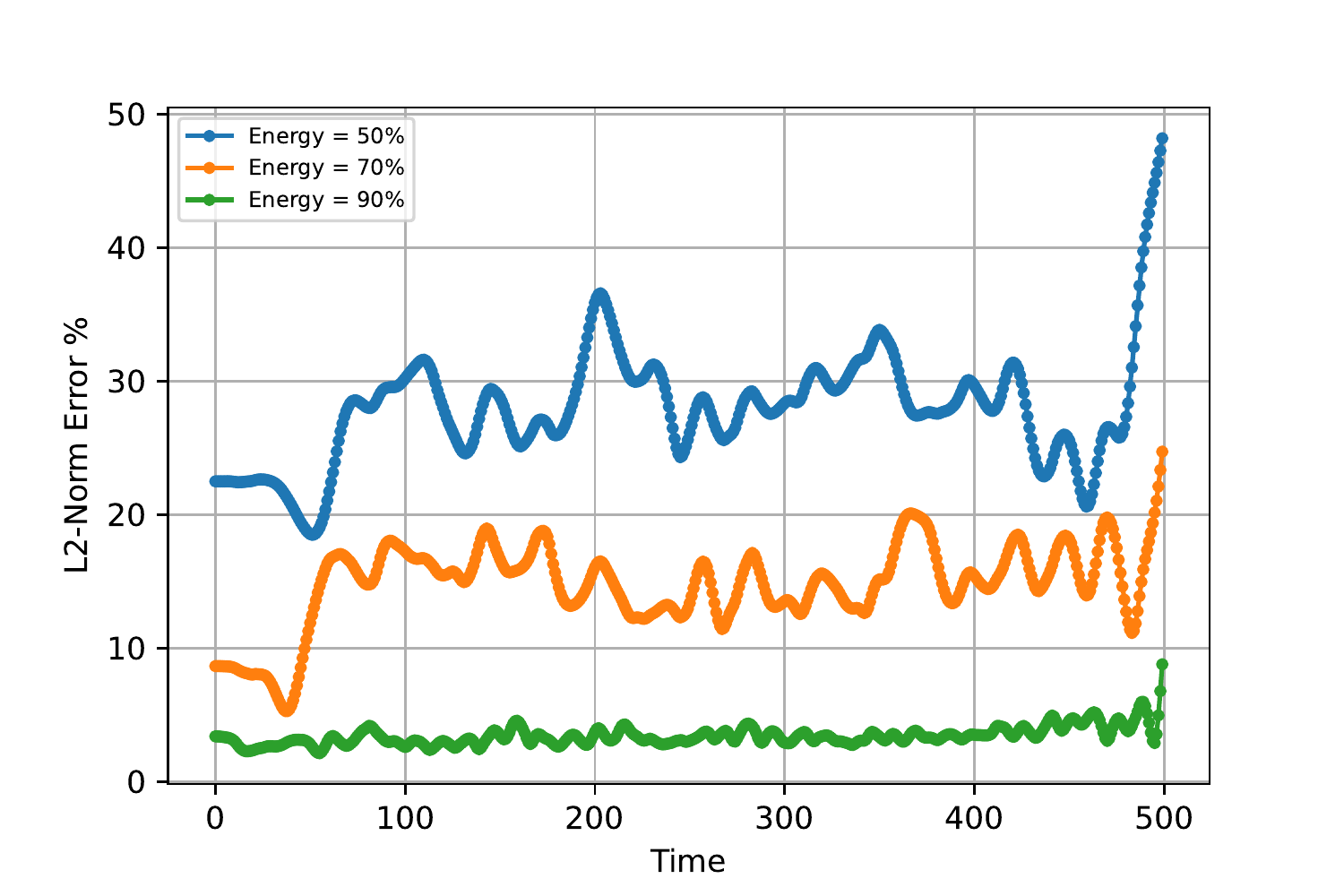}}\par 
\subfloat[Particle position $\widetilde{z}$]{\label{fig:Error-diffEng_zp}\includegraphics[width=.5\linewidth]{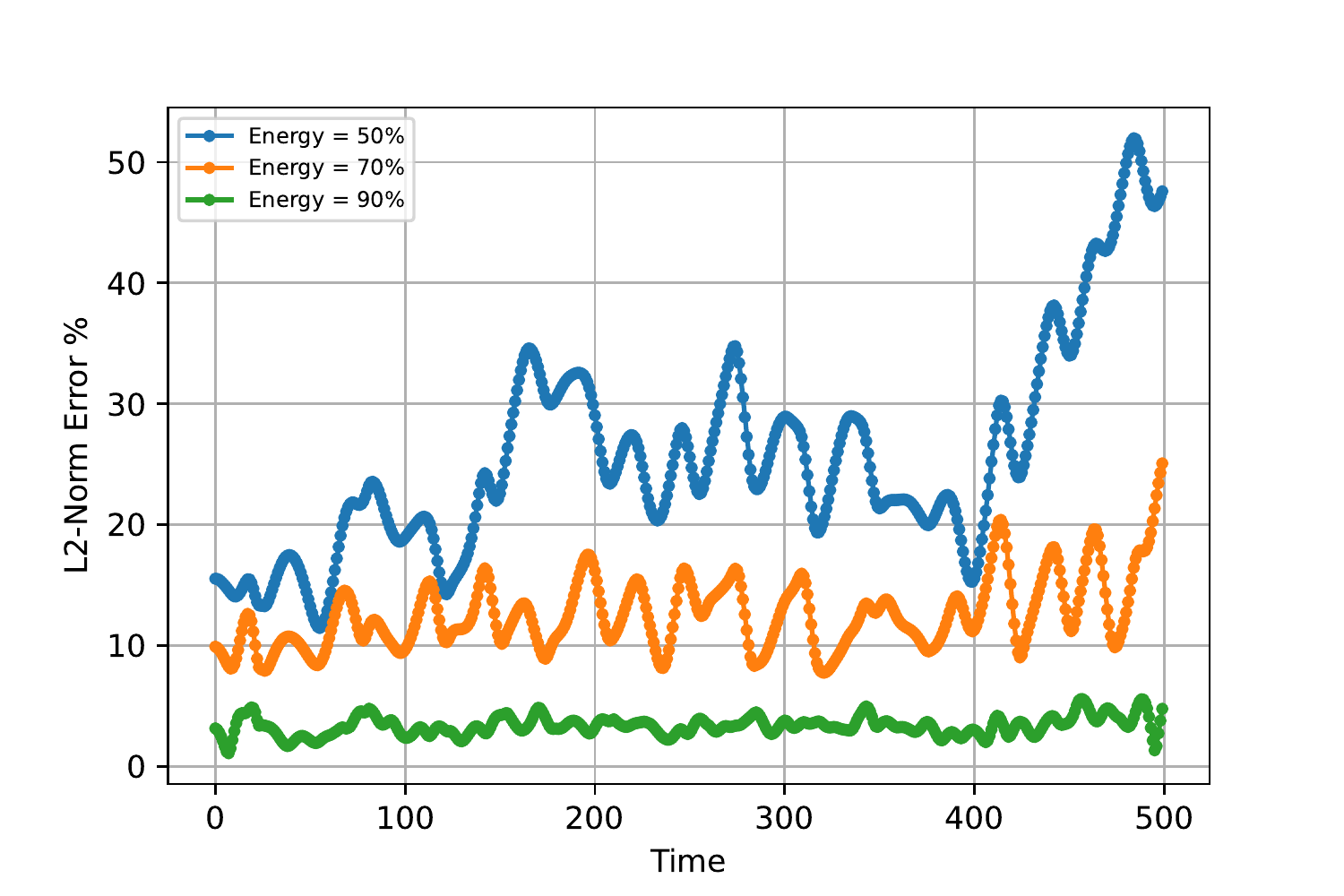}}
\caption{ROM validation - time reconstruction of the Lagrangian field: time evolution of the $L^2$-norm relative error (see eq. \eqref{eq:error2}) for different energy levels.} 
\label{fig:Error_in_time_lag}
\end{figure}

Also in this case, at the aim to provide a more quantitative comparison, we compute the time evolution of the $L^2$-norm relative error between FOM and ROM for the particle position:

\begin{equation}\label{eq:error2}
E_{\Phi}(t) = 100 \cdot \dfrac{||\Phi_h(t) - \Phi_r(t)||_{L^2(\Omega)}}{||{\Phi_h}(t)||_{L^2(\Omega)}},
\end{equation}
where $\Phi_h = \{\widetilde{x}, \widetilde{y}, \widetilde{z}\}$ is the field computed with the FOM and $\Phi_r = \{\widetilde{x}_r, \widetilde{y}_r, \widetilde{z}_r\}$ is the corresponding field computed with the ROM.
Again we consider three different values of $\delta$: 50\%, 70\% and 90\%. The results are reported \fig{fig:Error_in_time_lag}. 
As expected, 
the error is lower when a greater number POD modes is considered for the reconstruction. For $\delta = 50\%$, 
an error of diverging-like behaviour for the final time of the simulation, due to the lack of high order POD modes, is exhibited. On the other hand, for $\delta = 70\%$ such trend is mitigated and for $\delta = 90\%$ it completely disappears.
We report in \tab{tab:Part_Pos_error} the statistics of the $L^2$-norm relative error. We can observe that maximum, mean and minimum errors halves moving from 50\% to 70\% and decrease of about one order of magnitude when reaching the 90\%. 

\begin{table}[ht!]
\centering
 \begin{tabular}{||c |c |c |c||} 
 \hline
    Energy threshold & $\delta=50\%$ & $\delta=70\%$ & $\delta=90\%$ \\ [0.5ex] 
 \hline\hline
 Mean Error (\%) & 23.93 & 12 &  1.56 \\ \hline
 Max Error (\%) & 51.97 & 25.08 &  5.92 \\ \hline
 Min Error (\%) & 11.46 & 5.31 &  0.26 \\
 \hline
 \end{tabular}
 \caption{ROM validation - time reconstruction of the Lagrangian field: mean, maximum and minimum values of the $L^2$-norm relative error (see eq. \eqref{eq:error2}) by using the $90\%$ of the starting FOM database to train our ROM model for three different energy thresholds.}
 \label{tab:Part_Pos_error}
\end{table}

\begin{figure}[ht]
    \centering       \includegraphics[width=80mm,scale=0.5]{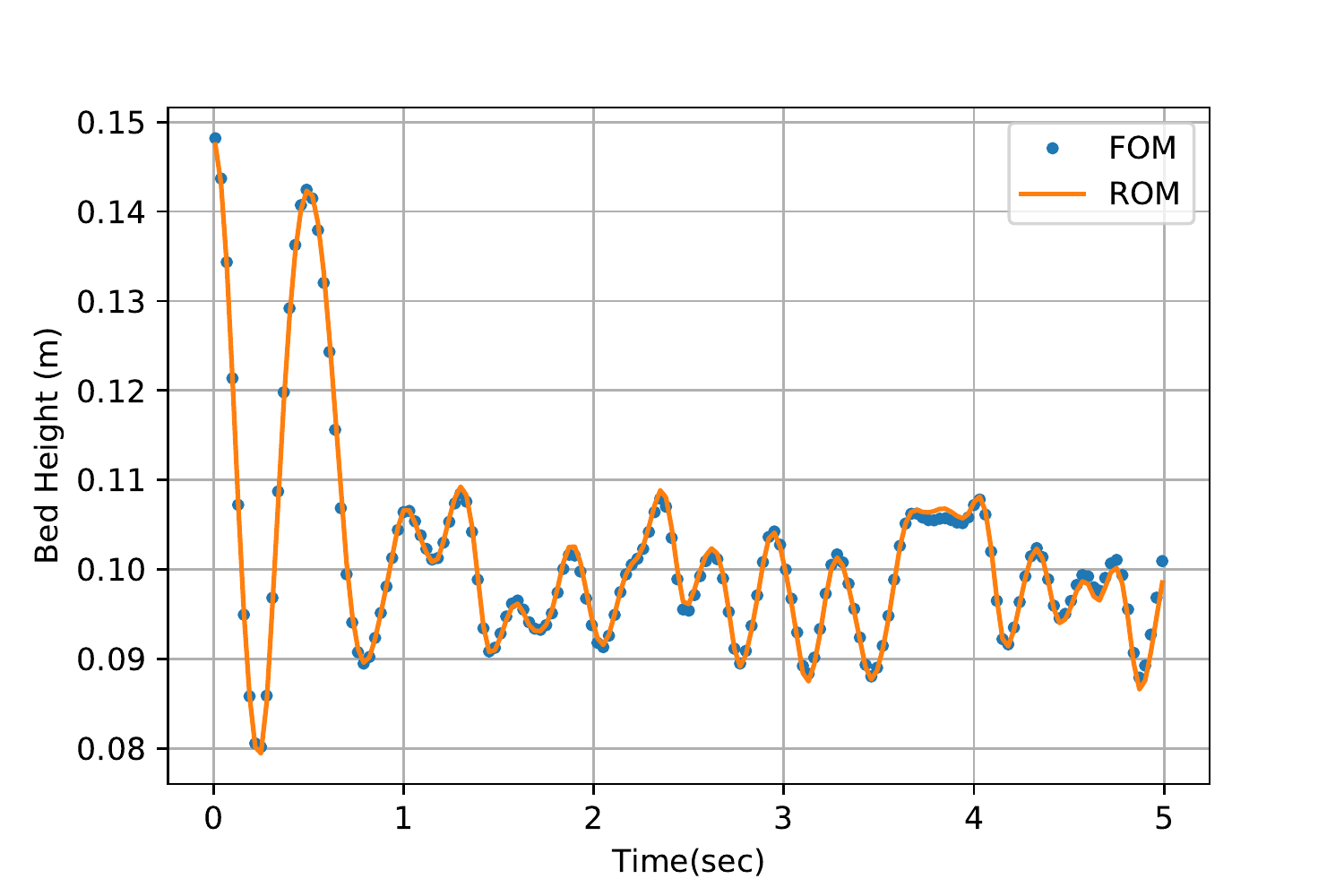}
        \caption{ROM validation - time reconstruction of the Lagrangian field: time evolution of the bed height (see eq. \eqref{eq:bed}) computed by FOM and ROM.} 
        \label{fig:bed_Height}
\end{figure}

For further investigation, we also compare the FOM and ROM predictions of the particle bed height $H_{bed}$ computed as:\\
\begin{equation}\label{eq:bed}
H_{bed}= \frac{\sum_{i=1}^{n_{p}} \widetilde{z}_i }{n_p}, \end{equation}
The results are shown in \fig{fig:bed_Height}. As can be noted, we obtain a very good agreement between the two solutions. 

Finally, we comment on the computational cost. We ran the FOM and ROM simulations on an 11th Gen Intel$(R)$ Core(TM) i7-11700 $@$ 2.50GHz 32GB RAM by using one processor.
The FOM simulation takes around 1.8\textrm{e}5 s, while the computation of reduced coefficients takes 0.11 s for $\epsilon$, 0.24 s for $\widetilde{\bm{x}}$ and 0.54 s for $\widetilde{\bm{u}}$. Therefore, we obtain a global speed-up of  $2\textrm{e}5$.


We can conclude that overall our ROM technique shows a very good performance both in terms
of efficiency and accuracy in the time reconstruction of the Eulerian and Lagrangian flow fields, provided that a
sufficient number of POD modes and training high-fidelity snapshots are considered.


\subsection{Parametrization with respect to the Stokes number $Stk$}\label{sec:Stokes_param}
\hfill \break
The Stokes number ${Stk}$, defined in eq. \eqref{Stk}, represents a crucial parameter of the model. So, after having
investigated the ability of our ROM approach to reconstruct the time evolution of Eulerian and Lagrangian fields, we build a parametric ROM with respect to $Stk$. In this context, it should be noted that we limit to consider the Eulerian phase. We are going to address a complete parametric framework including also the Lagrangian phase in a future work as reported in Sec. \ref{sec:conc}.

We choose a uniform sample distribution in the range $Stk \in [200, 300]$. To vary $Stk$ we modify the value of $\rho_p$. We consider 21 sampling points. For each value of $Stk$ in such set, a simulation is run for the entire time interval of interest, i.e. $(0, 5]$. Based on the results presented for the time reconstruction, the snapshots are collected every 0.02 s, for a total number of 10500 snapshots.
To train the ROM we consider the 90$\%$ of the database for each simulation, i.e. 450 snapshots, randomly chosen,  resulting in a total of 9450 snapshots. The remaining ones are considered as validation set. 
As already mentioned in the manuscript, two different PODI strategies are employed: the global PODI and the local PODI (see Sec. \ref{sec:ROM}). 

\begin{figure}[ht]
\centering
\subfloat[ Global PODI]{\label{fig:CumEnergy_global}\includegraphics[width=.52\linewidth]{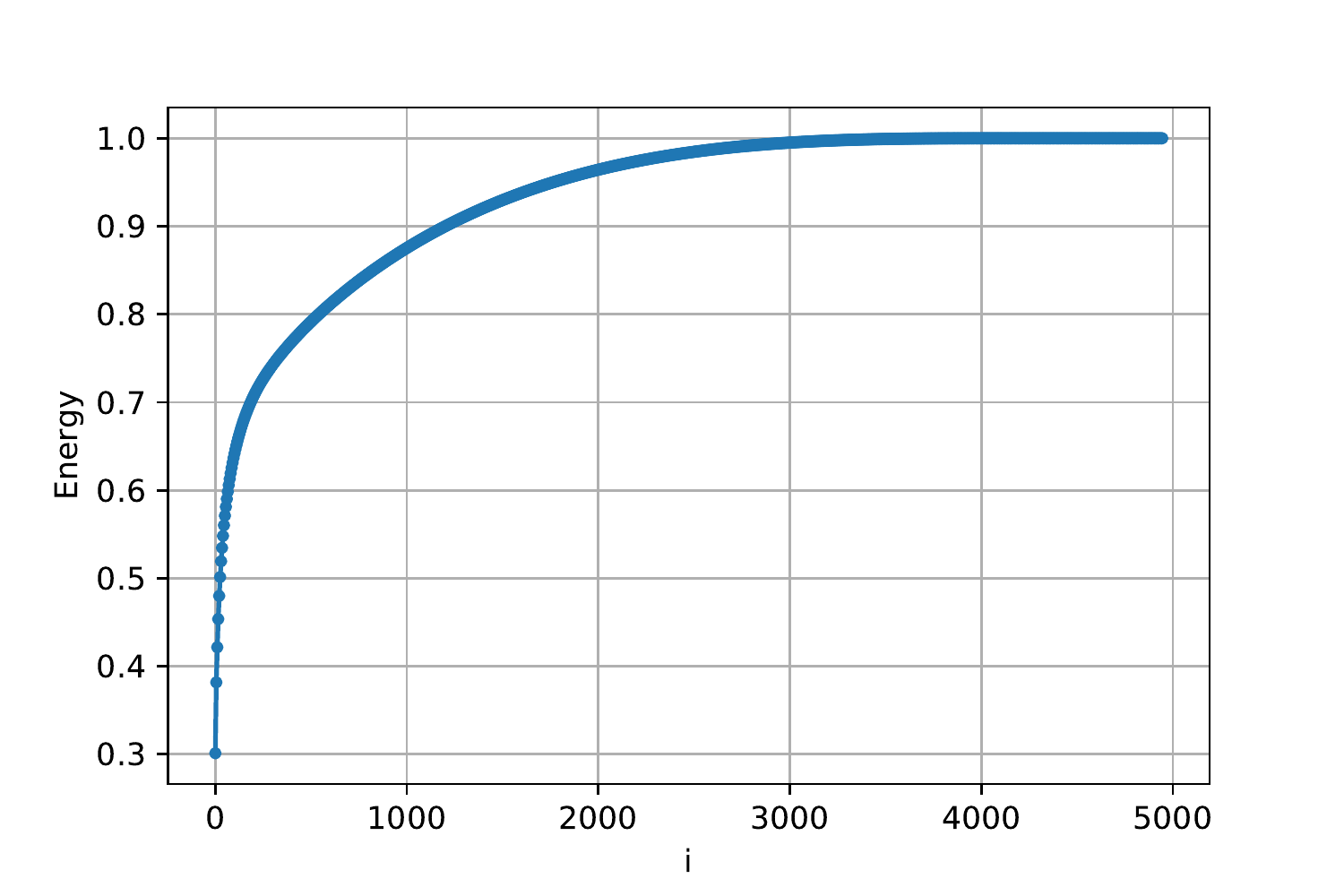}}
\subfloat[Local PODI]{\label{fig:CumEnergy_local}\includegraphics[width=.52\linewidth]{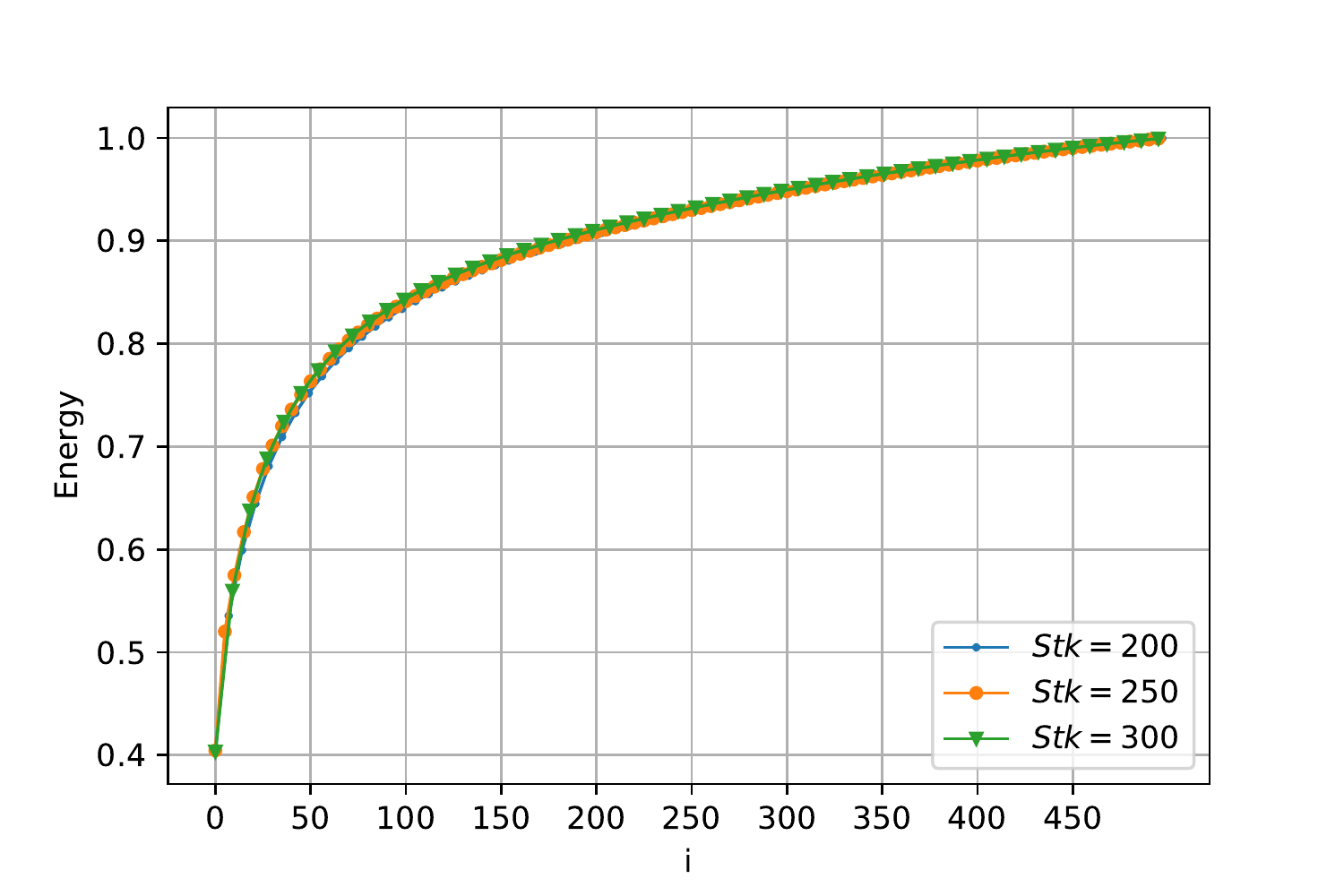}}
\caption{ROM validation - parametrization with respect to $Stk$: cumulative eigenvalues for the fluid volume fraction $\epsilon$ by global PODI (a) and local PODI (b). }
\label{fig:Energy_local_Global}
\end{figure}

The plot of the cumulative eigenvalues for the fluid volume fraction $\epsilon$ is shown in \fig{fig:Energy_local_Global}. For the global PODI approach (a), we have truncated the plot at 5000 modes, since the 99.99\% of the energy is already reached for 3200 modes. 
However, we oberve that at least 1500 modes are necessary to recover the 90\% of the cumulative energy. 
On the other hand, concerning the local PODI (b), we show the plot of the cumulative eigenvalues for the initial, mid and final value of $Stk$ in the training set, 200, 250 and 300 respectively. 
The comparison shows no significant difference between the three curves suggesting that the same number of modes needs to be considered to capture the same energy threshold. In particular, 150 modes can capture the 90$\%$ of the energy for the whole range of $Stk$, one order of magnitude in less than the global PODI.

\begin{figure}
    \centering
    \begin{subfigure}{0.7\textwidth}
        \includegraphics[width=\linewidth]{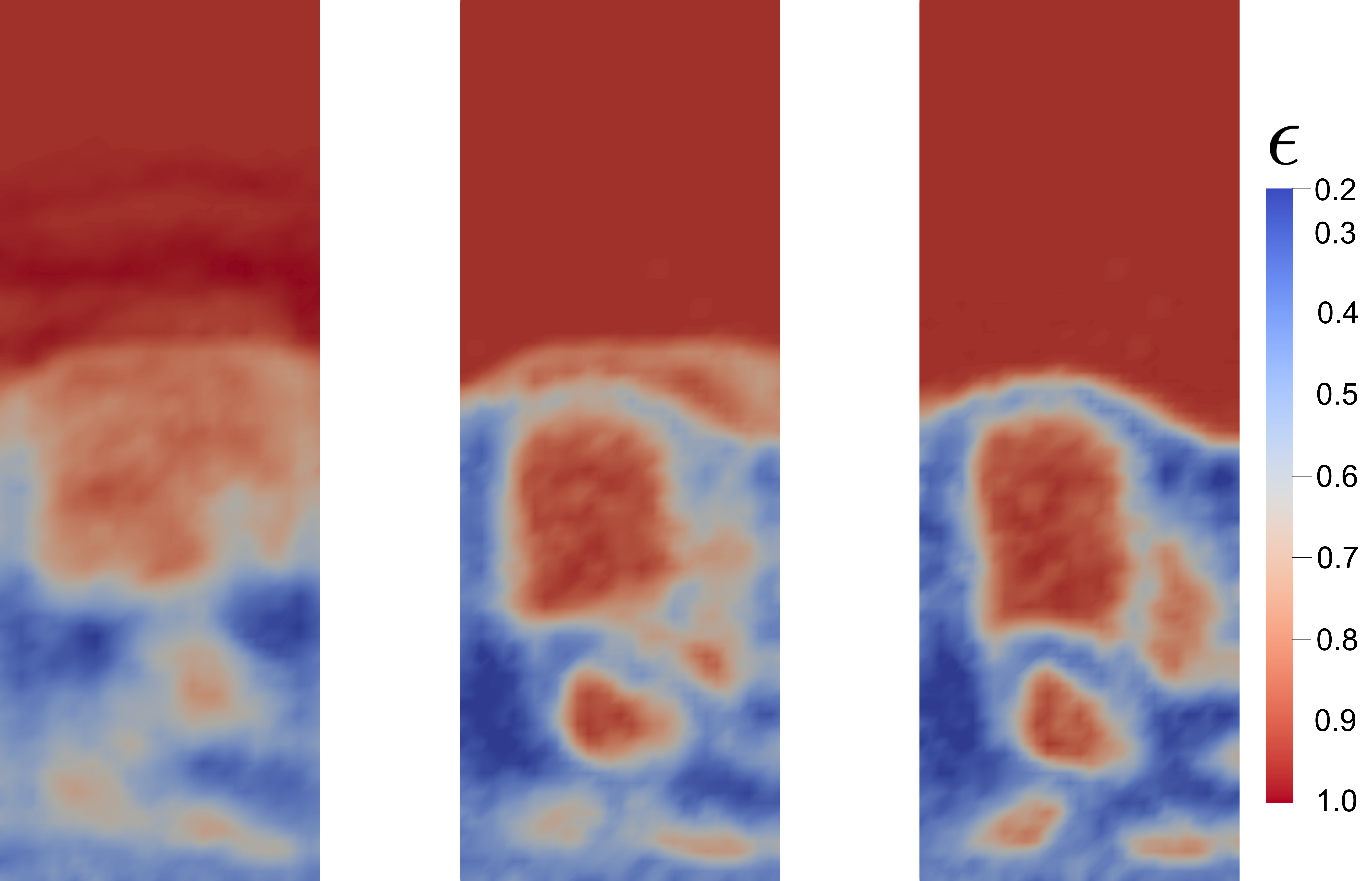}
        \label{fig:St227_t=1}
    \end{subfigure}
\medskip
    \begin{subfigure}{0.7\textwidth}
        \includegraphics[width=\linewidth]{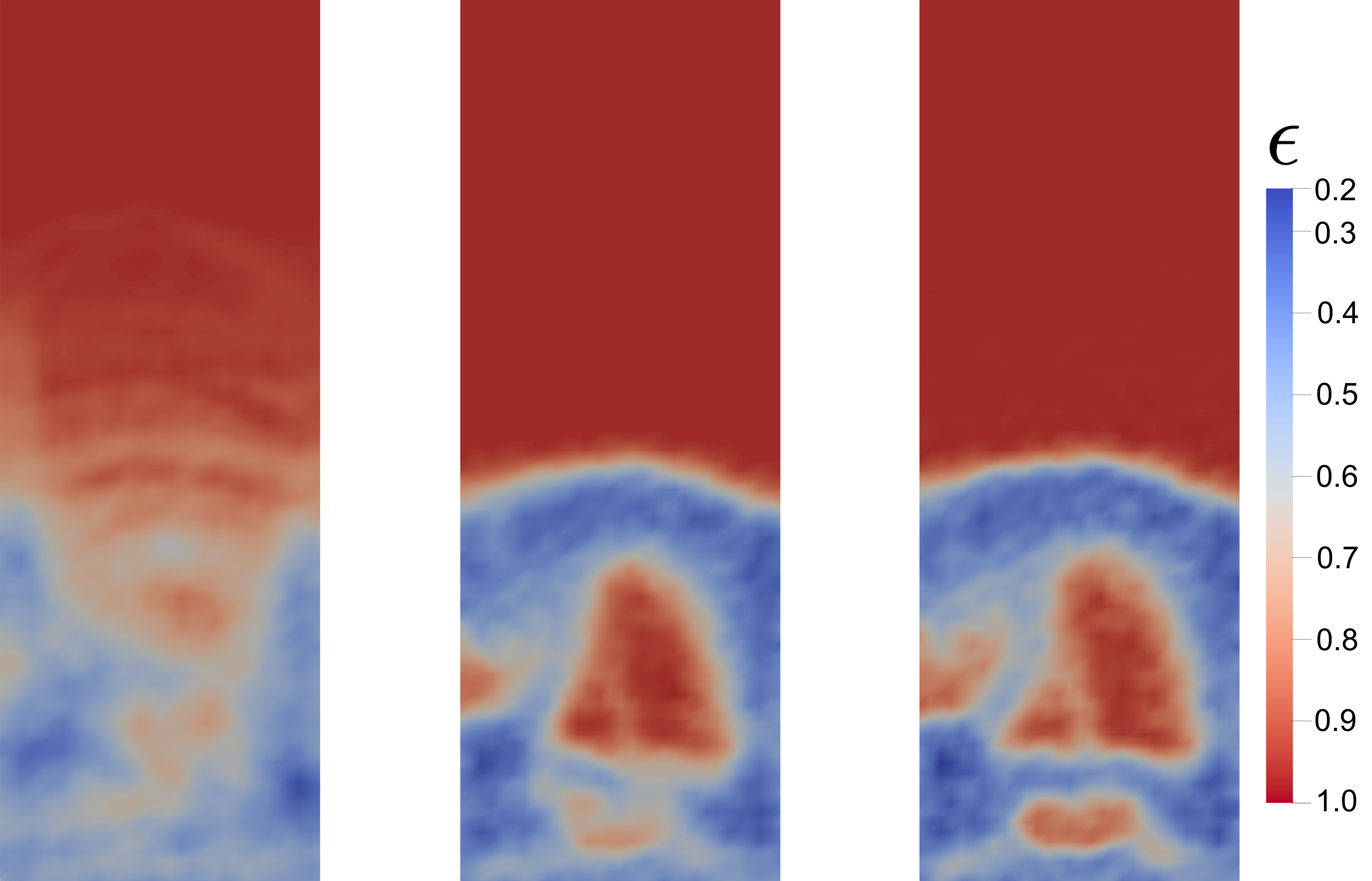}
        \label{fig:St227_t=2.5}
    \end{subfigure}
    \medskip
    \begin{subfigure}{0.7\textwidth}
        \includegraphics[width=\linewidth]{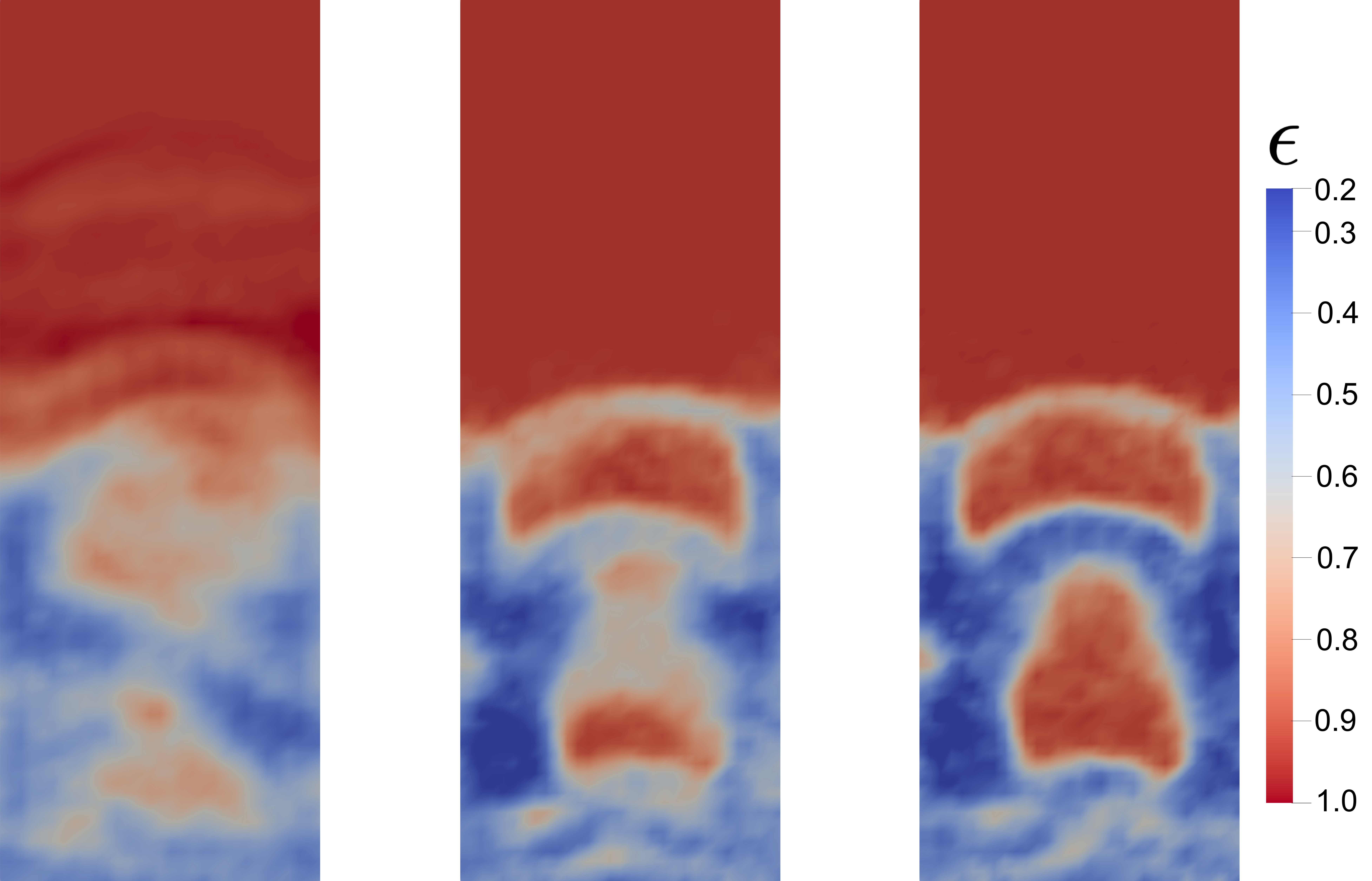}
        \label{fig:St227_t=4}
    \end{subfigure}    
\caption{ROM validation - parametrization with respect to $Stk$: comparison between FOM and ROM for $Stk =227$ for three different time instances: $t = 1$ s (first raw), $t = 2.5$ s (second raw) and $t = 4$ s (third raw). The first column represents the solution by global PODI, the second one the solution by local PODI and the third one the solution by FOM.} 
\label{fig:Vis_ParROM_St227}
\end{figure}

\begin{figure}
    \centering
    \begin{subfigure}{0.7\textwidth}
        \includegraphics[width=\linewidth]{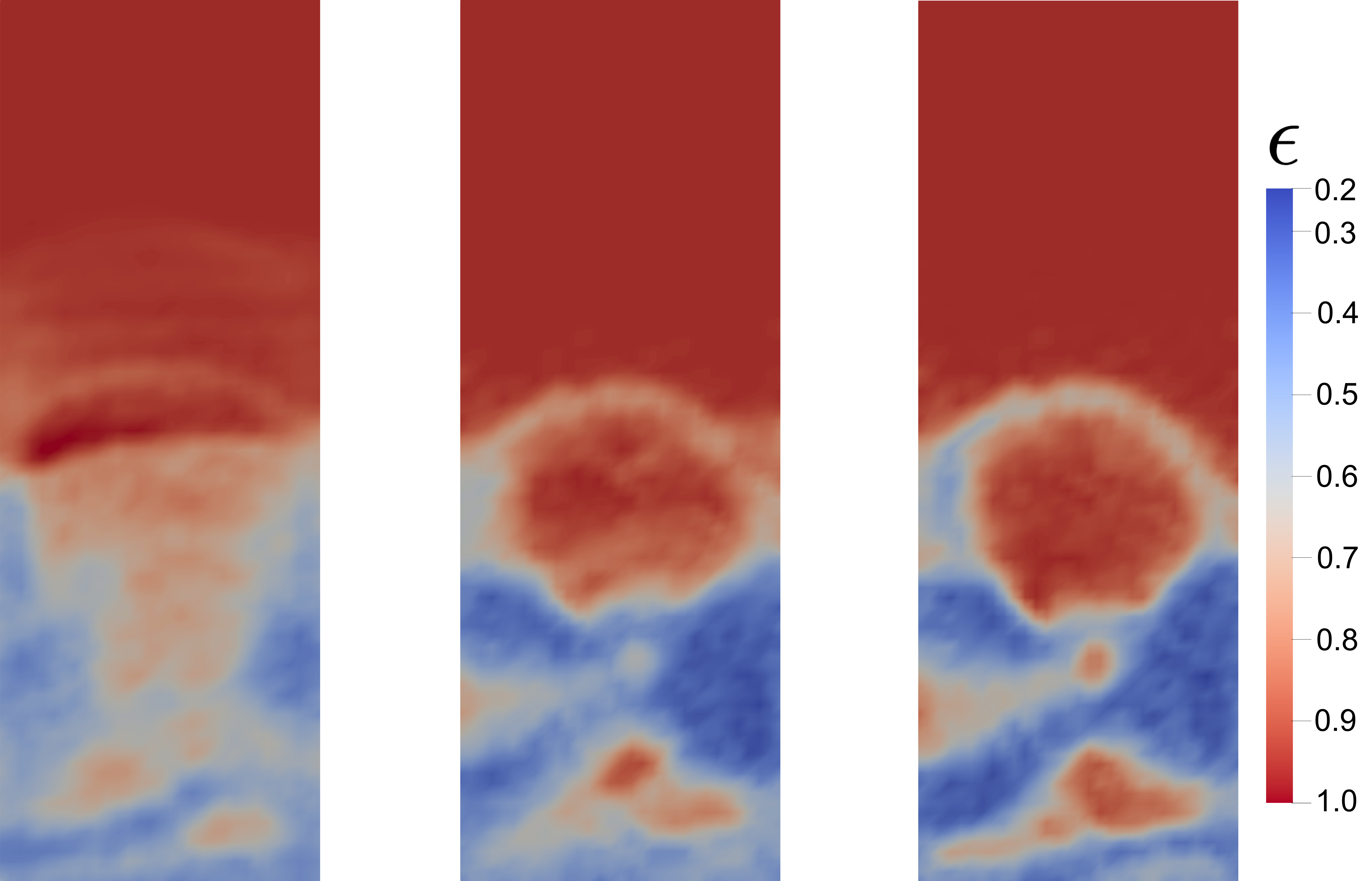}
        \label{fig:St277_t=1}
    \end{subfigure}
\medskip
    \begin{subfigure}{0.7\textwidth}
        \includegraphics[width=\linewidth]{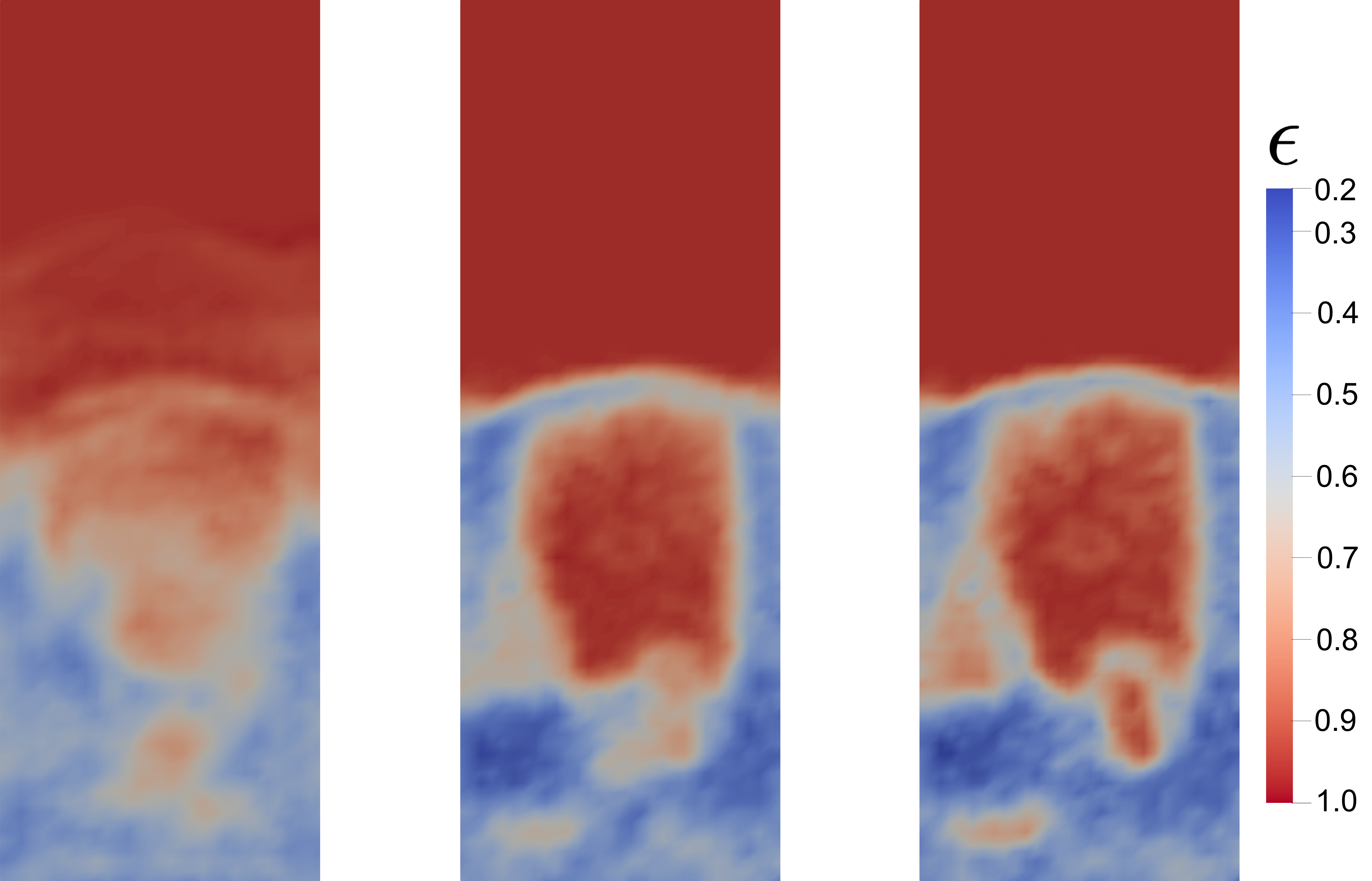}
        \label{fig:St277_t=2.5}
    \end{subfigure}
    \medskip
    \begin{subfigure}{0.7\textwidth}
        \includegraphics[width=\linewidth]{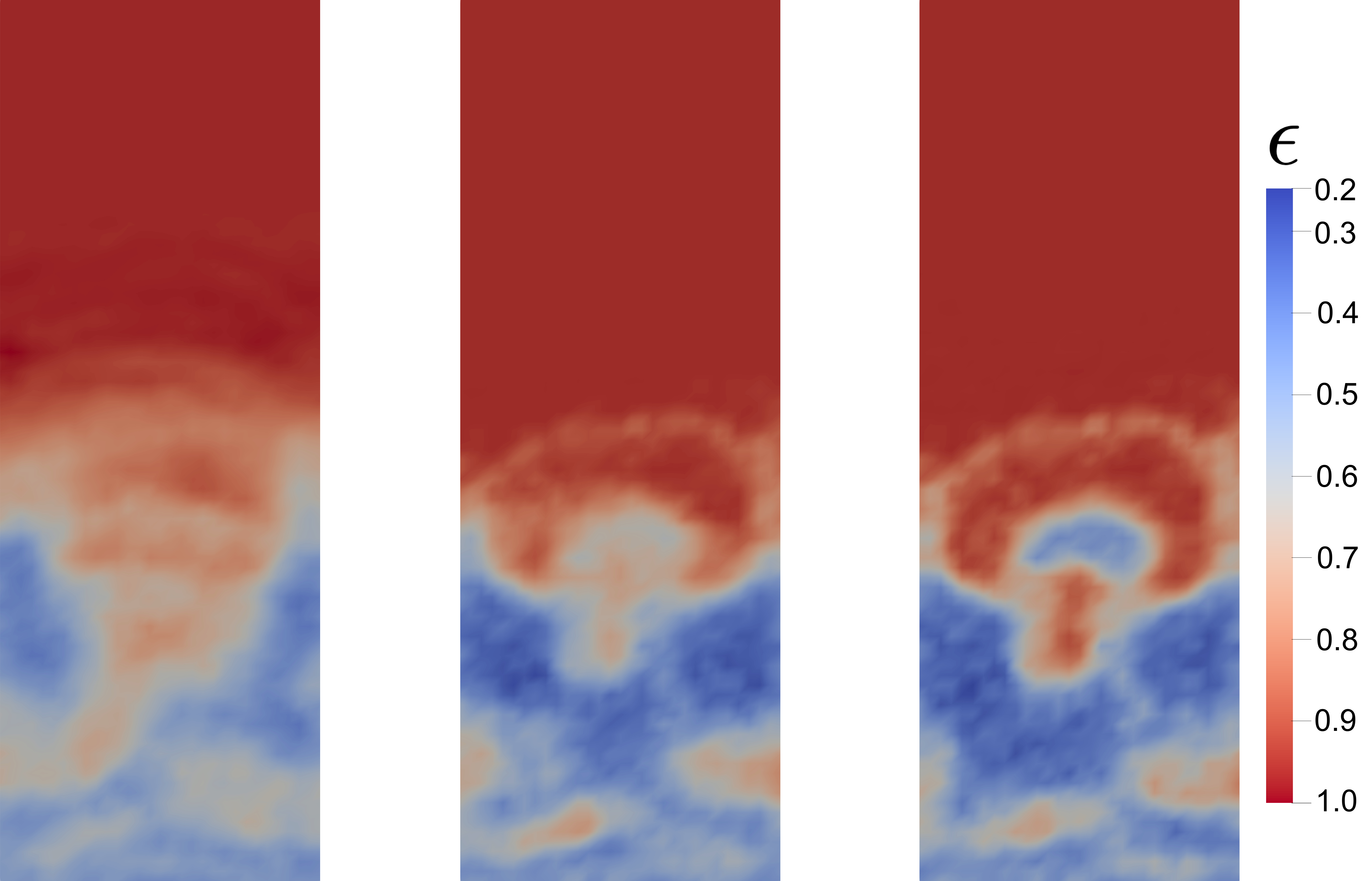}
        \label{fig:St277_t=4}
    \end{subfigure}    
\caption{ROM validation - parametrization with respect to $Stk$: comparison between FOM and ROM for $Stk =277$ for three different time instances: $t = 1$ s (first raw), $t = 2.5$ s (second raw) and $t = 4$ s (third raw). The first column represents the solution by global PODI, the second one the solution by local PODI and the third one the solution by FOM.} 
\label{fig:Vis_ParROM_St277}
\end{figure}

We take $Stk = 227$ and $Stk = 277$ as testing points to
evaluate the performance of the parametrized ROM. A comparison between the two reconstructed solutions and the corresponding FOM is shown in \fig{fig:Vis_ParROM_St227} and \fig{fig:Vis_ParROM_St277} for $t = 1$ s, $t = 2.5$ s and $t = 4$ s. 
As one can see, in both cases, the global PODI is barely able to reconstruct the main patterns of flow field. The solution is affected by some spurious oscillations. The situation changes using the local PODI which can capture more details and partially damps the unphysical oscillations. 

\begin{figure}[ht]
\centering
\subfloat[ $Stk$ = 227]{\label{fig:Stk227_error}\includegraphics[width=.52\linewidth]{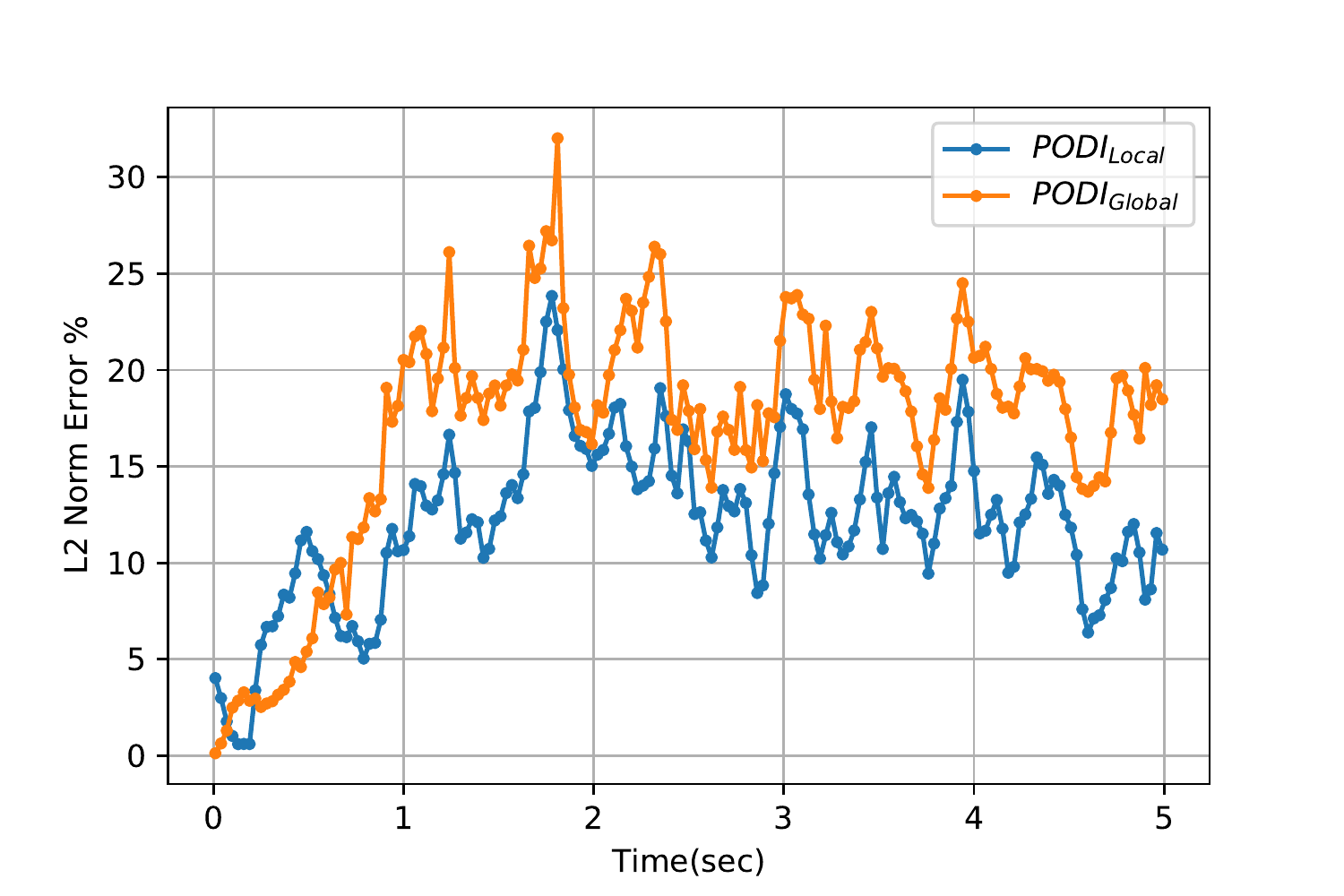}}
\subfloat[$Stk$ = 277]{\label{fig:Stk277_error}\includegraphics[width=.52\linewidth]{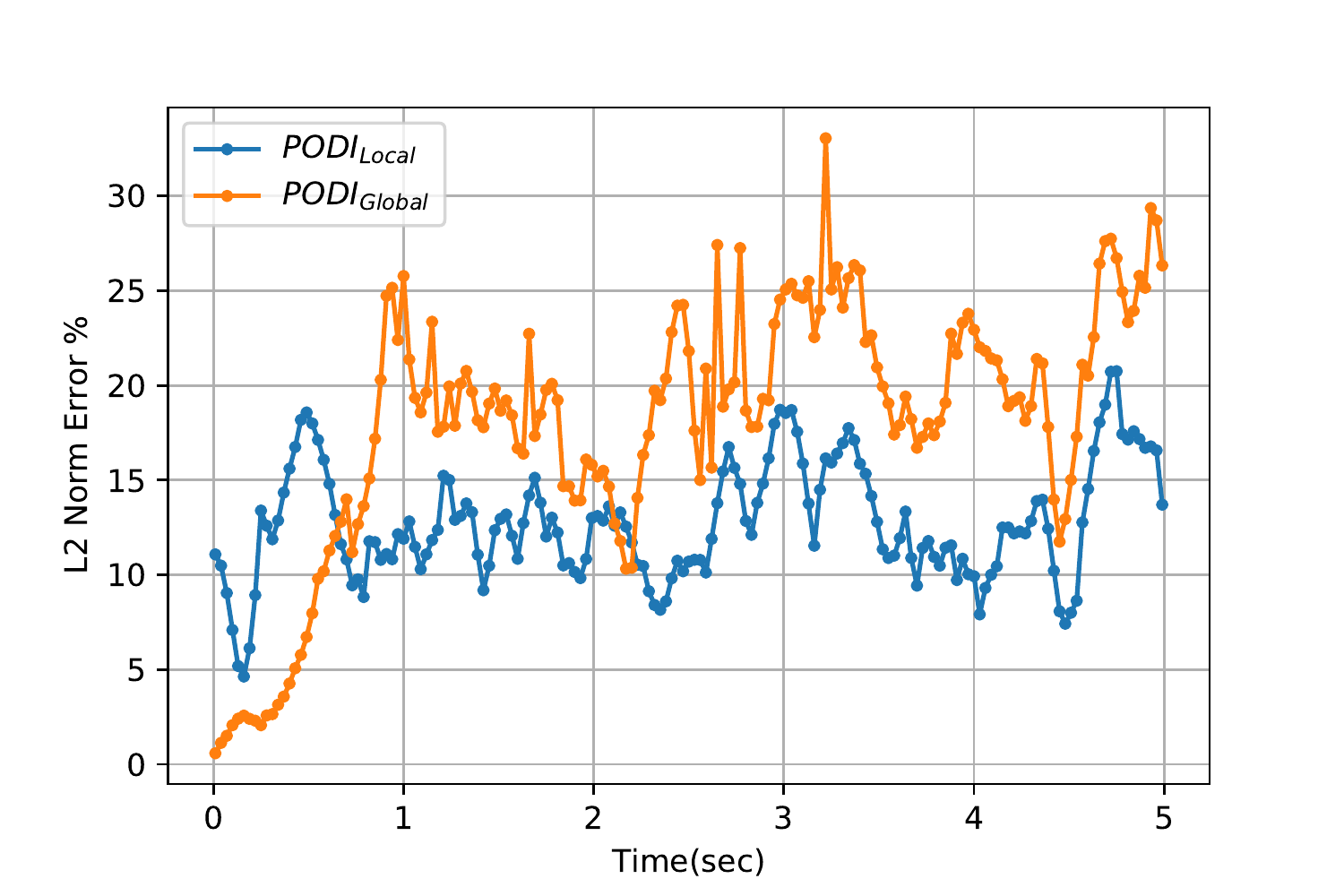}}
\caption{ROM validation - parametrization with respect to $Stk$: time evolution of the $L^2$-norm relative error (eq. \eqref{eq:error1}) for the two testing points. }
\label{fig:error_time_parametric}
\end{figure}

To introduce a more quantitative comparison, 
the time history of the $L^2$-norm relative error between FOM and ROM (eq. \eqref{eq:error1}) has been computed for both $Stk = 227$ and $Stk = 277$. The plot is shown in \fig{fig:error_time_parametric} whilst \tab{tab:Parametric_error} reports the values of the maximum, minimum and mean error. From \fig{fig:error_time_parametric}, we note that, except for the initial time of the simulation, the local PODI provides an error lower than the global one. The maximum error decreases of about two times when one adopts the local PODI. Also the mean error decreases passing from 17-18\% to 12\%. 




Beyond the capability of local PODI to capture better the system dynamics, it is also much cheaper in terms of CPU time of about one order of magnitude. Indeed the online phase associated the local PODI takes around 9 s, while the global PODI requires 121 s. 

So we can conclude that the local PODI basically performs better than the global one in a parametric framework both in terms of efficiency and accuracy. 


\begin{table}[ht!]
\centering
\begin{tabular}{||c|c|c|c|c||}
\hline 
$Stk$ &  \multicolumn{2}{c|}{$227$}  &  \multicolumn{2}{c||}{$277$} \\
\hline\hline
ROM approach &  Local PODI & Global PODI & Local PODI & Global PODI \\
\hline
Mean Error (\%) &  12.2 & 18.1 & 12.1 & 17.2 \\ \hline
Max Error (\%) & 19 & 33 & 23 & 32 \\\hline
Min Error (\%) &4.3 & 0.6 & 0.4 & 0.1 \\\hline
\end{tabular}
 \caption{ROM validation - parametrization with respect to $Stk$: mean, maximum and minimum values of the $L^2$-norm relative error (eq. \eqref{eq:error1}) for the two testing points $Stk = 227$ and $Stk = 277$ by local and global PODI.}
 \label{tab:Parametric_error}
\end{table}

\newpage
\section{Concluding remarks}\label{sec:conc}

This work presents a non-intrusive data-driven ROM based on a PODI approach for fast and reliable CFD-DEM simulations.  

Unlike the previous works \cite{shuo, shuo2, mori}, we choose a Finite Volume method because of its computational efficiency making this study particularly appealing both for academic and industrial applications where multi-phase systems are involved.
We assessed our ROM approach through a classical benchmark adopted for the validation of CFD-DEM solvers: a fluidized bed two-phase flow system \cite{fernandes, tsu, heat}. 
We found that our ROM can capture the unsteady flow features with a good accuracy both for Eulerian and Lagrangian variables. 
We also performed a parametric study with respect to the Stokes number for the Eulerian phase. In this context, we have compared two different strategies: the global PODI and the local PODI. 




Moving forward, we will take care to improve the ROM reconstruction of the CFD-DEM simulations both in terms of efficiency and accuracy, especially in a parametric framework. At this aim, we are going to adopt Machine Learning (ML)-based techniques for the detection of the non-linear behaviour exhibited by the FOM. 
In particular, we plan to use Autoencoders as an alternative to POD that may capture, more efficiently, features or patterns in the high-fidelity model results \cite{lee2020model, fresca2021comprehensive} and Feed Forward Neural Networks (instead of RBFs) to map the reduced coefficients in the parameter space \cite{shah2022finite, hesthaven2018non}. 

\section{acknowledgments}\label{sec:akw}

We acknowledge the support provided by the European Research Council Executive Agency by the Consolidator Grant project AROMA-CFD "Advanced Reduced Order Methods with Applications in Computational Fluid Dynamics" - GA 681447, H2020-ERC CoG 2015 AROMA-CFD, the European Union’s Horizon 2020 research and innovation program under the Marie Skłodowska-Curie Actions, grant agreement 872442 (ARIA), PON “Research and Innovation on Green related
issues” FSE REACT-EU 2021 project, PRIN NA FROM-PDEs project, INdAM-GNCS 2019-
2021 projects and SISSA-Dompè project “Development of a CFD model for INNOJET VENTILUS V1000 granulator”. 

\newpage
\bibliographystyle{plain}
\bibliography{mybib}

\end{document}